\documentclass[a4paper,11pt]{article}
\usepackage[utf8]{inputenc}

\pdfoutput=1

\usepackage{jcappub}

\usepackage[english]{babel}
\usepackage{units}
\usepackage{siunitx}
\usepackage{graphicx}
\usepackage{amsmath}
\usepackage{amssymb}
\usepackage{dsfont}
\usepackage{subfigure}
\usepackage{lineno}
\usepackage{multirow}
\usepackage{tikz}
\usepackage{pgf}
\usepackage{xspace}
\usepackage{bbold}

\usepackage{natbib}
\usepackage{graphicx}

\definecolor{dark-gray}{gray}{0.3}

\begin{document}

\title{Reconstructing non-repeating radio pulses with Information Field Theory}
\author[a,b]{C.~Welling}
\author[c,d]{P.~Frank}
\author[c,d]{T.~Enßlin}
\author[a,b]{A.~Nelles}

\affiliation[a]{DESY, Platanenalle 6, Zeuthen, Germany}
\affiliation[b]{Erlangen Center for Astroparticle Physics, Friedrich-Alexander-Universit\"at Erlangen-N\"urnberg, Erwin-Rommel-Str.\ 1, 91058 Erlangen, Germany}
\affiliation[c]{Max Planck Institute for Astrophysics, Karl-Schwarzschildstraße 1, 85748 Garching, Germany}
\affiliation[d]{Ludwig-Maximilians-Universit{\"a}t M{\"u}nchen, Geschwister-Scholl-Platz 1, 80539, M{\"u}nchen, Germany}

\emailAdd{christoph.welling@desy.de}

\abstract{
    Particle showers in dielectric media produce radio signals which are used for the detection of both ultra-high energy cosmic rays and neutrinos with energies above a few PeV. The amplitude, polarization, and spectrum of these short, broadband radio pulses allow us to draw conclusions about the primary particles that caused them, as well as the mechanics of shower development and radio emission. However, confidently reconstructing the radio signals can pose a challenge, as they are often obscured by background noise. Information Field Theory offers a robust approach to this challenge by using Bayesian inference to calculate the most likely radio signal, given the recorded data. In this paper, we describe the application of Information Field Theory to radio signals from particle showers in both air and ice and demonstrate how accurately pulse parameters can be obtained from noisy data.
}

\maketitle

\section{Introduction}

The origin of the most energetic cosmic rays is still not conclusively identified. One approach to this problem could be \emph{multi-messenger astrophysics}, where several types of cosmic particles are used to identify the sources of these ultra-high energy cosmic rays (UHECRs). Whatever sources produce UHECRs likely also emit high-energy neutrinos \cite{Halzen:2016gng, murase_waxman}, which are not deflected on their way to Earth. Additionally, the most energetic cosmic rays can interact with the cosmic microwave background to create high-energy neutrinos via the GZK effect \cite{gzk_greisen, gzk_zatsepin, gzk_kuzmin}. Finding these \emph{cosmogenic neutrinos} would also allow us to draw conclusions about the origins of UHECRs \cite{vanVliet:2019nse}.

Considering this connection, it is not surprising that both cosmic ray and neutrino detectors face the same main problem at the highest energies: The steeply falling flux requires large effective areas, which lead to the construction of cosmic ray observatories that cover thousands of square kilometers \cite{auger_prime, telescope_array} and neutrino detectors with fiducial volumes on the cubic kilometer scale \cite{ice_cube_overview, km3net_letter_of_intent, baikal_overview}. This is possible by not detecting UHECRs directly, but the showers they produce when interacting with the Earth's atmosphere, in the case of cosmic rays, or bodies of water or ice, in the case of neutrinos. One way of detecting these showers is to measure the radio signals they produce.

\subsection{Detecting radio signals from particle showers}
The emission of radio signals by particle showers in dielectric media has been proposed as early as the 1960s \cite{askaryan, jelley}. At the time, technical limitations prevented its practical use for particle detectors, but advances in digital technology have turned this technique into a powerful method to detect high-energy particle showers. While it is used successfully for the detection of cosmic rays in several experiments (e.g.\ \cite{aera_inclined_showers, lofar, codalema}), the use for neutrino detection in ice is just about to transition from prototype experiments \cite{ARIANNAdesign, ARA} to discovery-scale detectors \cite{rnog_whitepaper, gen2collaboration2020icecubegen2}.

For a particle shower to emit strong radio signals, two conditions have to be met: First, there needs to be a separation of positive and negative charges in the shower front and the signals produced over the length of the shower profile need to overlap coherently, e.g.\ \cite{Huege:2016veh}. The separation of charges can be caused by two primary mechanisms: The \emph{Askaryan effect} is the formation of a negative net charge in the shower front because positrons annihilate with electrons in the surrounding matter and Compton scattering accelerates electrons to become part of the shower. The other effect, called \emph{geomagnetic emission} is the separation of charges by the Lorentz force from the geomagnetic field, which creates a positive and negative pole in the shower front. Because of its relatively high density, in ice the Askaryan effect is dominant. For air showers, geomagnetic emission is usually stronger, but the Askaryan effect has a significant contribution. It is possible to distinguish between both by the the different polarization of their radio signals \cite{lofar_polarization}. With Askaryan emission, the polarization is radial towards the shower axis while geomagnetic emission produces polarization in the direction of the Lorentz force.

The second required condition is similar to the Cherenkov effect. If an observer is positioned at the Cherenkov angle ($\sim 1^\circ$ in air, $\sim 54^\circ$ in ice) all radio emission produced during the shower development reaches it at the same time, leading to constructive interference. This means that the radio signal is strong on a cone around the shower axis. Compared to light in the visible spectrum, however, the much longer radio waves can maintain coherence more easily, so that even a few degrees off the Cherenkov angle, the radio signal is still strong enough to be detected, especially at lower frequencies. 

After the radio signal has been emitted, it needs to propagate to the detector, during which it can be altered by propagation effects. For air showers, this is very straightforward: Air is almost perfectly transparent for radio waves, so attenuation is negligible and the only modification to consider is a decrease in the electric field amplitude $|\vec{E}| \sim \frac{1}{R}$ with distance $R$ of the observer from the shower maximum due to a widening of the illuminated area.
In ice, the propagation of the radio signal from the emission region to the observer is much more complicated. The index of refraction of glacial ice changes with depth, which causes the ray path of the radio signal to be bent downward, instead of propagating in a straight line. Horizontal propagation of the radio signal has been observed as well \cite{ara_ice_profile, horizontal_propagation_antarctica, horizontal_propagation_summit}, it seems, however, a secondary effect. The radio signal can also be reflected at the ice surface. Because of this, it is possible for multiple radio pulses from one shower to reach the same antenna. The attenuation length of radio signals in ice is on the order of \SI{1}{km} \cite{ross_ice_shelf_ice_properties, summit_ice_properties, southpole_ice_properties}, meaning that the absorption of the signal on the way to the detector has to be taken into account. The attenuation length is also frequency-dependent, meaning that it can change the shape of the frequency spectrum of the signal.

\subsection{Properties of radio signals from particle showers}
The radio signal from a particle manifests as a short, linearly polarized pulse with a duration of the order of a few nanoseconds. In general, the amplitude of the radio signal is proportional to the shower energy $E_\mathrm{shower}$, making it possible to reconstruct the energy of the primary particle from the radio signal. In practice, however, it is often easier to use the energy fluence of the radio signal
\begin{equation}
    \phi_E = c \cdot \epsilon_0 \cdot \int \vec{E}^2(t) dt
    \label{eq:energy_fluence}
\end{equation}
for energy reconstruction, which scales with shower energy as $\phi_E \sim E_\mathrm{shower}^2$  \cite{PhysRevD.93.122005}.

\begin{figure}
    \centering
    \includegraphics[width=.9\textwidth]{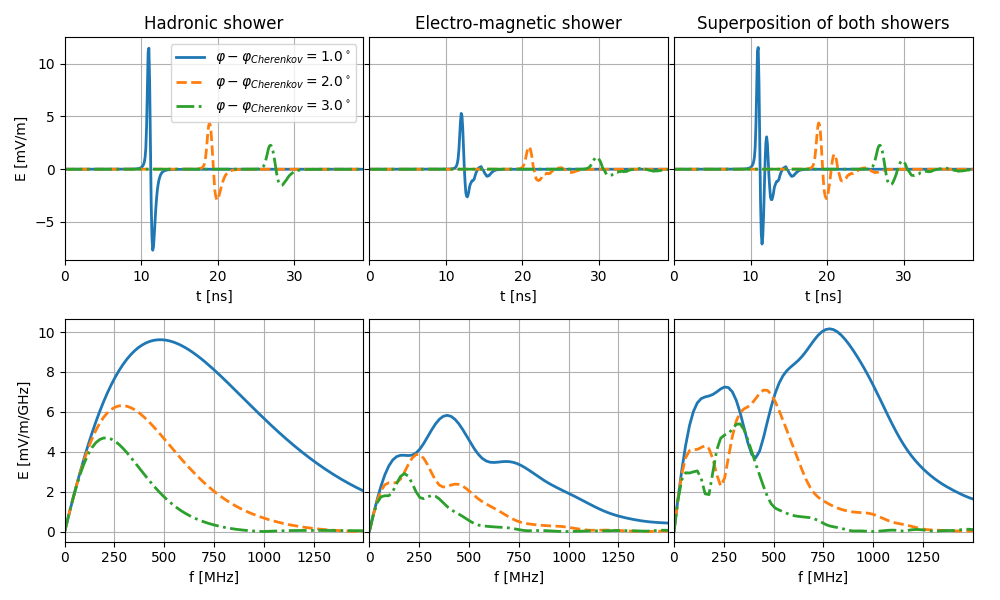}
    \caption{Examples of the radio signals from a particle shower in ice with a deposited energy of \SI{1e18}{eV} observed at a distance of \SI{2}{km} in the time domain (top) and frequency domain (bottom). Left: Radio signals from hadronic showers. Middle: Radio signals from electro-magnetic showers. Right: Superposition of the signals from a hadronic and an electro-magnetic shower. For better readability, the different pulses have been shifted in time. Simulations were performed with the \emph{NuRadioMC} software framework \cite{NuRadioMC} using the \emph{ARZ} emission model \cite{Alvarez_Mu_iz_2010}.}
    \label{fig:signal_examples}
\end{figure}

The amplitude also depends on the angle between the shower axis and the direction in which the signal is emitted, the so-called viewing angle. This is shown for in-ice showers in Fig.~\ref{fig:signal_examples}: As the difference between viewing angle and Cherenkov angle increases, the radio waves emitted during the development of the shower lose coherence and the overall signal becomes weaker. Looking at the frequency spectrum, this decrease in amplitude is not uniform over the whole bandwidth, but shorter wavelengths lose coherence more quickly and are suppressed. This means that, unless the viewing angle is very close to the Cherenkov angle, the radio signal is usually stronger at lower frequencies. This viewing angle dependence means that the shape of the frequency spectrum is an important property for the reconstruction of the shower energy, as it can be used as a proxy to correct for the loss of coherence if the radio signal is observed off of the Cherenkov angle. This technique has already been shown to work in the case of air showers \cite{Welling:2019scz} and is expected to be the preferred method for neutrino detection \cite{rnog_whitepaper}.

In the case of air showers, the frequency spectrum roughly follows an exponential function \cite{NuRadioReco}, for showers in ice, its shape is more complicated, as shown in Fig.~\ref{fig:signal_examples}. While a hadronic shower still produces a radio signal with a rather regular spectrum, a high-energy electro-magnetic shower is affected by the LPM effect \cite{lpm1, lpm2}, making it longer and less smooth \cite{PhysRevD.82.074017}. This means that there are effectively multiple smaller shower maxima, whose radio emissions interfere with each other, leading to a more irregular shape of the frequency spectrum. Even if the shower energy is too small for the LPM effect to occur, the charged-current interaction of an electron neutrino in ice will produce two particle showers: One hadronic shower from the nuclear recoil and one electro-magnetic shower from the electron produced in the interaction. The radio emissions from both showers interfere, which also leads to a rather complicated shape of the frequency spectrum. This makes it difficult to find a general analytic parameterization of the frequency spectrum, like it exists for air showers.

Because of these relationships between shower properties and the radio signal, calculating the electric field is an important step in the event reconstruction process of many radio detection experiments. In this paper, we present a new method to reconstruct the electric field from noisy antenna measurements, the software for which is freely available as part of the \emph{NuRadioReco} software package \cite{NuRadioReco}. How to best use the results from this method will depend on the specific detector setup and the goal of the analysis. One may also imagine using the presented method for event identification, however, we focus on reconstruction and provide a small outlook on possible applications in Sec.~\ref{sec:implications}.

\subsection{Reconstructing the electric field from noisy data}
\label{sec:classic_reconstruction}

For the reconstruction of the electric field $E(t)$ of the radio pulse from the  recorded time-dependent voltage traces (waveforms) $U(t)$, it is useful to transform them into the frequency domain via Fast Fourier Transformation (FFT) to get the frequency spectrum $\mathcal{V}_i(f)$. The electric field is also expressed in the frequency domain by the two components $\mathcal{E}^\theta(f)$ and $\mathcal{E}^\phi(f)$, which describe the polarizations in the $\vec{e}_\theta$ and $\vec{e}_\phi$ directions, respectively\footnote{Since the detector is in the far field,  the electric field in $\vec{e}_r$ direction can be assumed to be 0.}. If the incoming direction $(\theta, \phi)$ of the radio signal is known, the antenna response of channel $i$ to the $\vec{e}_\theta$ and $\vec{e}_\phi$ components of the electric field can be expressed by the vector effective length $\mathcal{H}_i^\theta(f, \theta, \phi)$ and $\mathcal{H}_i^\phi(f, \theta, \phi)$. The result is a system of linear equations, connecting the electric field to the waveform recorded by each channel:
\begin{equation}
    \begin{pmatrix} \mathcal{V}_1(f) \\ \mathcal{V}_2(f) \\ ...\\ \mathcal{V}_n(f)\end{pmatrix} = 
    \begin{pmatrix} \mathcal{H}_1^\theta (f)& \mathcal{H}_1^\phi (f)\\ \mathcal{H}_2^\theta (f) & \mathcal{H}_2^\phi (f)\\ ... \\ \mathcal{H}_n^\theta (f)& \mathcal{H}_n^\phi (f)\end{pmatrix} 
    \begin{pmatrix} \mathcal{E}^\theta(f) \\ \mathcal{E}^\phi(f)\end{pmatrix} \
    \label{eq:H_full}
\end{equation}

If data from enough channels ($n=2$) is available, in principle, solving this system of equations yields a reconstruction of the electric field (e.g.\  \cite{Abreu:2011fb}). In practice, however, the recorded voltage waveforms are always contaminated by a certain amount of noise, so the resulting electric field will be contaminated, too. Especially, if the system of equations is over-determined ($n>2$) an exact solution for (\ref{eq:H_full}) does not exist. In this case, the electric field is typically reconstructed by minimizing the sum of the squared errors on $\mathcal{V}_i(f)$ for each frequency bin (e.g.\ \cite{NuRadioReco}).

\begin{figure}
    \centering
    \includegraphics[width=.9\textwidth]{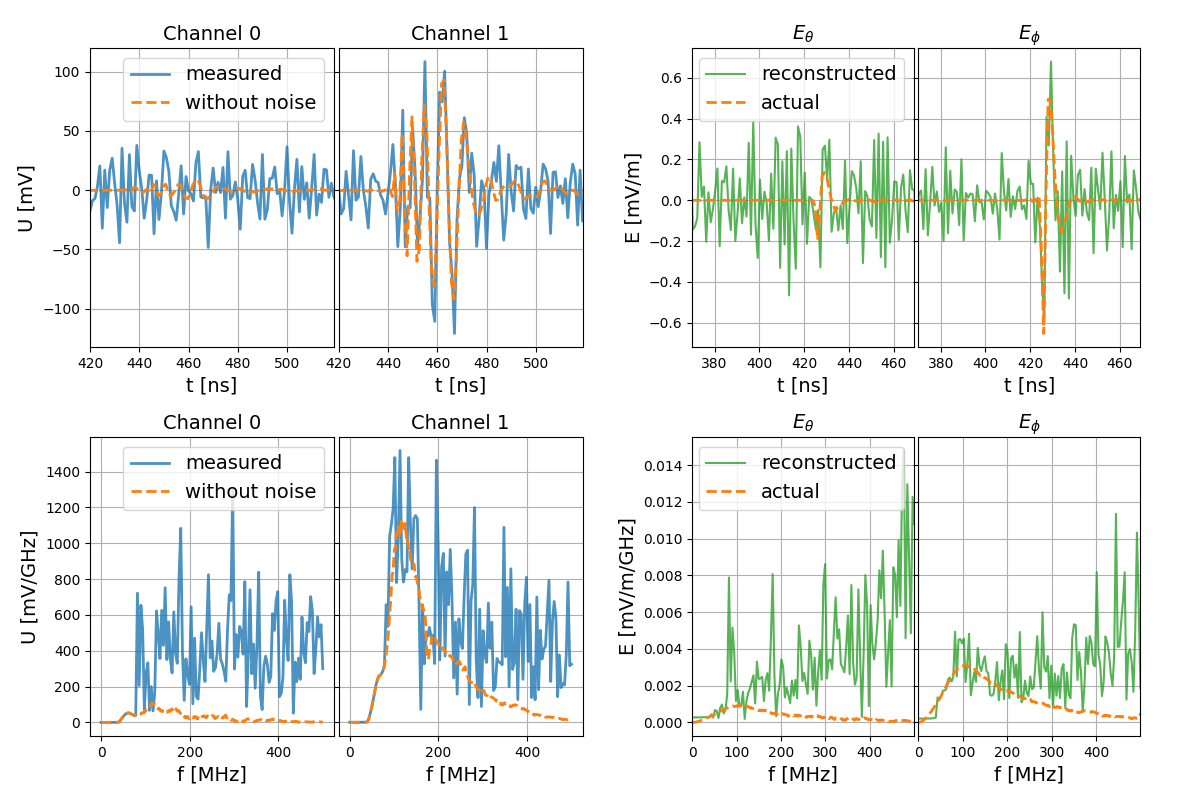}
    \caption{Example for the electric field reconstruction from noisy data by solving (\ref{eq:H_full}). Left: Signals with (blue solid) and without noise (orange dotted) in the time-domian (top) and frequency domain (bottom). Right: Reconstructed electric field (green solid) in the time domain (top) and frequency domain (bottom) compared to the true noiseless electric field (orange dotted).}
    \label{fig:classic_efield_rec}
\end{figure}

This method works well in many situations, but it has some shortcomings, especially for low signal-to-noise ratios (SNR): Fig.~\ref{fig:classic_efield_rec} shows an example, in which the radio signal from an air shower is reconstructed from the waveforms measured by two logarithmic-periodic dipole antennas (LPDAs). Both are pointing directly upwards and are angled \SI{90}{^\circ} relative to each other, so that one is sensitive to the north-south and the other to the east-west component of the electric field. If one of the components of the electric field is small, as is the case for the $\vec{e}_\theta$ component in this example, it is overestimated, because noise is interpreted as coming from the radio signal. Especially if none of the antennas has a high sensitivity to one of the electric field components, errors in the reconstruction of the signal polarization can be large. For example, this would be the case for very inclined air showers, for which $\vec{e}_\theta$ is almost vertical, if it is detected by the antennas are only sensitive to the horizontal electric fields.

Another issue the the shape of the frequency spectrum. It is usually not possible to construct a broadband antenna that has a uniform sensitivity over the entire frequency band. This can cause the reconstruction to overestimate the electric field at frequencies with low sensitivity, as any noise there is interpreted as being caused by the radio signal and not by noise. In Fig.~\ref{fig:classic_efield_rec}, this is what causes the reconstructed electric field spectrum to rise with higher frequencies.

For these reasons, more advanced reconstruction methods are needed if one wants to use the electric field polarization and especially the shape of the spectrum for event reconstruction. In the case of air showers, a \emph{forward folding} method has previously been developed, which fits a parameterization of the frequency spectrum to the data \cite{NuRadioReco,Welling:2019scz}. Such  parameterizations also exist for particle showers in ice, so the same approach is in principle possible for neutrino detectors. There are, however, caveats. While the shape of the frequency spectrum is relatively simple for hadronic showers and low-energy electromagnetic showers, the LPM effect causes the shower to become elongated and more irregular at high energies, which in turn makes the shape of the spectrum more irregular (see Fig.\ref{fig:signal_examples}). On top of that, charged-current interactions of $\nu_e$ cause both a hadronic and an electro-magnetic shower, whose signals can interfere and cause a complicated spectral shape. This makes \emph{forward folding} with a single signal parameterization a challenging approach for neutrino detectors.

Additionally, it seems prudent to avoid too much reliance on parameterizations derived from theoretical calculations. No neutrino interaction in ice has so far been identified by its radio signal, which would confirm that the spectrum of the electric field matches the predictions. Also, building bias-free confidence in these models when having detected a neutrino, will not be possible, if the reconstruction of the electric field is already using model-predictions.
On top of that, propagation effects could alter the radio signal on its way from the source to the detector e.g.\ \cite{horizontal_propagation_summit,horizontal_propagation_antarctica,Jordan:2019bqu,parabolic_equations}. These will be difficult to incorporate into an analytic parameterizartion, especially if the ice properties are not well known.

Therefore, we need a method that can reconstruct the electric field from the noisy waveforms recorded by the antennas, while also making only minimal, and well-founded, assumptions about what the result should look like. \emph{Information Field Theory} can provide such a method.


\subsection{Information Field Theory (IFT)} 

IFT \cite{ift1, ift2} is a statistical field theory developed to perform probabilistic reasoning about quantities (i.e.\ fields) that are defined over a continuous space. Specifically it extends information theory to square-integrable functions and allows for a probabilistic inference of such functions from noisy and incomplete observational data. IFT has successfully been applied in the past to various astrophysical data analysis tasks such as for example radio interferometry \cite{2019A&A...627A.134A, arras2020variable}, galactic tomography \cite{2020A&A...639A.138L, 2019A&A...631A..32L}, and galactic all sky imaging \cite{2020A&A...633A.150H}. As fields are infinite dimensional quantities whereas the data is always finite dimensional, such inference problems are ill-posed and therefore a prior model is inevitable.

In this work a prior model for the electric field is defined by means of a generative process. Specifically
\begin{equation}
    \mathcal{E}  = s\left(\xi\right) \ ,
    \label{eq:ift_generative}
\end{equation}
where all elements of $\xi$ follow the same probability distribution, called standard distribution, and therefore are independent and identically distributed random variables. Note that every continuous prior process can be expressed as a generative process by means of inverse transform sampling. In this work the standard distribution is set to be a zero mean Gaussian distribution with unit variance. This makes sampling from the prior distribution (see Fig.~\ref{fig:priors}) very straightforward as a sample from the prior process is defined as the result of applying the generative mapping $s$ to a sample from the standard distribution.

Combining the prior model with the likelihood of observing the measured voltage $U$ given the electric field $\mathcal{E}$ gives rise to the joint distribution of $U$ and $\xi$ via the product rule for probabilities as
\begin{equation}
    \mathcal{P}\left(\xi, U\right) = \mathcal{P}\left(U | \mathcal{E} = s\left(\xi\right)\right) \ \mathcal{P}\left(\xi\right) = \mathcal{P}\left(U | \mathcal{E} = s\left(\xi\right)\right) \ \mathcal{N}\left(\xi|0 , \mathds{1}\right) \ ,
    \label{eq:joint_prob}
\end{equation}
which is proportional to the posterior distribution $\mathcal{P}\left(\xi | U\right)$ up to a normalization factor independent of $\xi$.

For all but the simplest combinations of prior processes and likelihoods, integration w.r.t.\ the posterior is not possible analytically and therefore posterior expectation values cannot be computed directly and have to be approximated numerically. In this work we employ an approximation method based on variational inference, specifically we use Metric Gaussian Variational Inference (MGVI) \cite{knollmuller2019metric}. MGVI approximates the posterior distribution as a Gaussian distribution with the additional constraint that the covariance is set to be the inverse of the Fisher-Metric, evaluated at the mean of the Gaussian distribution. Approximation is achieved by numerical minimization of the forward Kullback–Leibler divergence between the posterior distribution and the approximation w.r.t.\ the mean. This provides an accurate estimation of the first moment in terms of the mean, and provides a lower bound to the second moment of the posterior distribution. The posterior is accessed via sampling from the approximate distribution and posterior expectation values are approximated by means of sample averages. Further details of the numerical approximation can be found in \cite{knollmuller2019metric}.


\section{Using IFT for the reconstruction of electric-field pulses}

Using IFT to reconstruct the electric field of a radio pulse means, stated in IFT terms, constructing the probability $\mathcal{P}(\mathcal{E} | U)$, which states the probability of the electric field, given the voltages measured by the radio antennas and prior assumptions. For this, we need a model for the prior distribution of the electric field, a model of the detector response, which connects the electric field to the measured voltages, and of the noise.

We do this for two distinct, but similar use cases: The radio detection of neutrinos and of cosmic rays.

\subsection{Use case: In-ice neutrino detection}

\begin{figure}
    \centering
    \includegraphics[width=0.52\columnwidth]{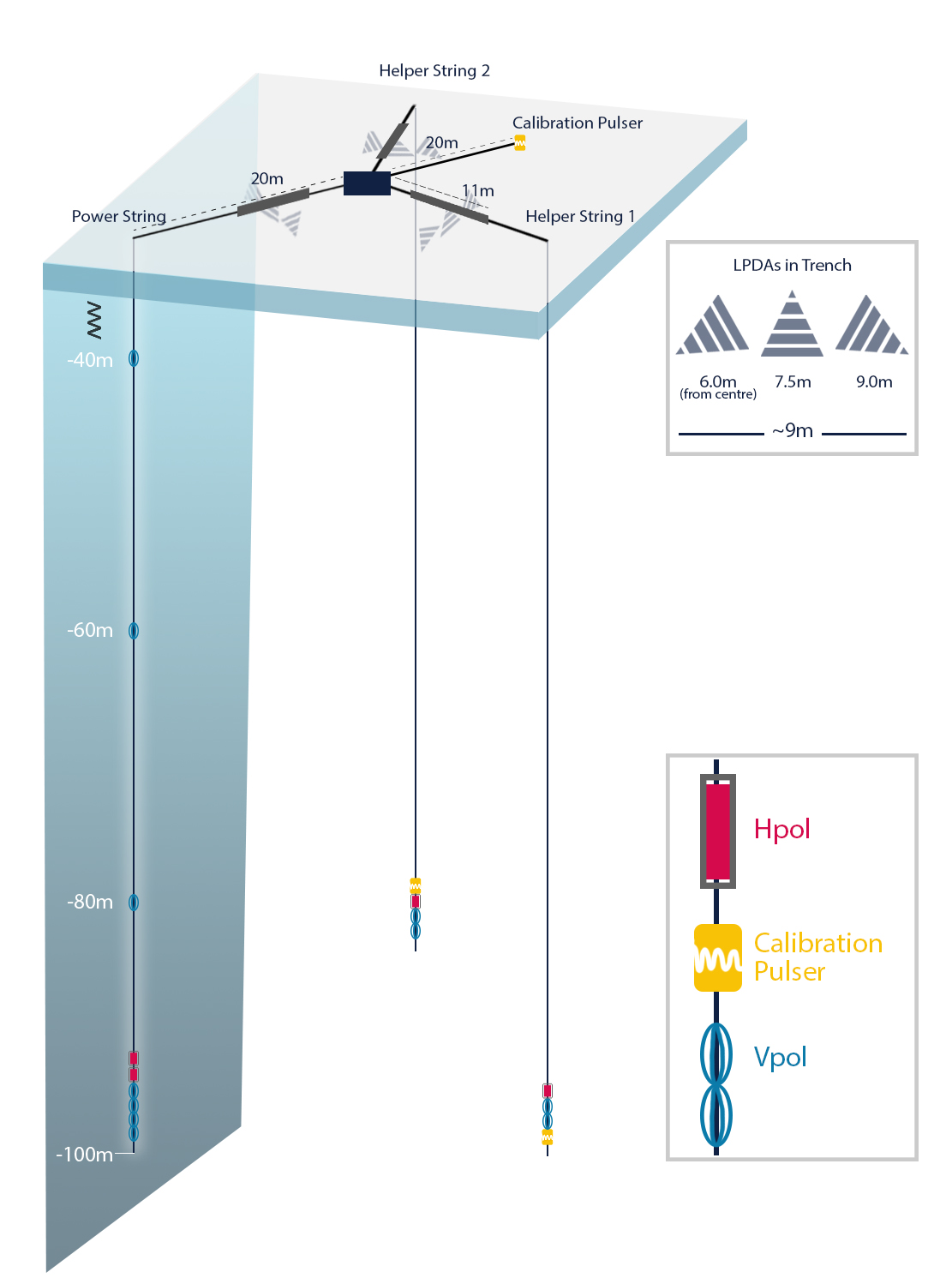}
    \includegraphics[width=.47\columnwidth]{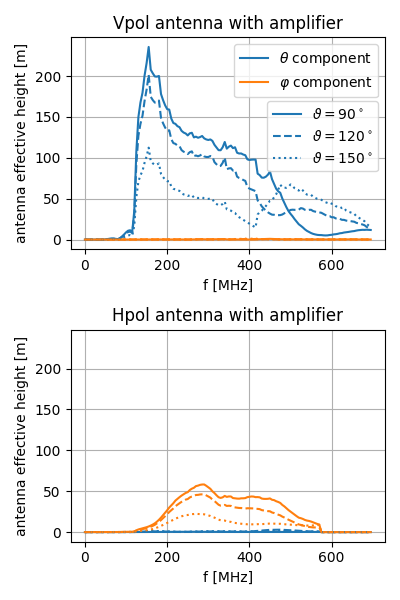}
    \caption{Left: One of the 35 stations that make up the RNO-G detector. The figure shows the station design as used for this analysis. RNO-G is still under construction and the station design may change. In particular the performance of the Hpol antennas will likely improve over what is shown in this article. For details, see \cite{rnog_whitepaper}, from where this figure is taken. Right: Response 
    of the amplifier and the antenna of the Vpol (top) and Hpol (bottom) channels as used in this analysis.}
    \label{fig:rno_station}

\end{figure}
For the case of in-ice detection of neutrinos, we show a simulation study of the Radio Neutrino Observatory in Greenland (RNO-G) \cite{rnog_whitepaper}. RNO-G will consist of a total of 35 independent detector stations, placed on a grid with a spacing of \SI{1.5}{km} at Summit Station, Greenland. Because of the large spacing, most neutrinos will be detected by only one station. Therefore, we consider only a single RNO-G station for this simulation study.

The layout of one RNO-G station is shown in Fig.~\ref{fig:rno_station}. It can be separated into a \emph{shallow} and a \emph{deep} component. The \emph{shallow} component consists of 3 arms with 3 logarithmic-periodic dipole antennas (LPDAs) on each of them. 
While they have excellent broadband sensitivity, their shallow depth makes it less likely that a radio signal from within the ice can reach them, so for most events, measurements from the LPDAs will not be available.
The \emph{deep} component consists of 3 holes going down to a depth of \SI{100}{m} below the surface, in which cable strings with radio antennas on them are placed. One of the strings, called \emph{power string}, holds a total of 9 antennas. Four of them are vertically polarized (Vpol) antennas spaced \SI{1}{m} apart, which form a phased array trigger at the very bottom of the string. Three more Vpol antennas are positioned with a spacing of \SI{20}{m} further up the string. Additionally, two horizontally polarized (Hpol) antennas are placed directly above the phased array. The other two so-called \emph{helper strings} are more sparsely instrumented, with two Vpol and one Hpol antennas at a depth of about \SI{100}{m}.

The design of the radio antennas is limited by the width of the boreholes they have to fit in. As shown in Fig.~\ref{fig:rno_station}, for this reason the Vpol antennas have a higher gain than the Hpol antennas and a maximum gain at lower frequencies. Together with the fact that the radio signals detected with RNO-G tend to be more strongly polarized in the vertical direction, this means that  the signal detected by the Hpol channels will often be hidden below the noise level, while it is only visible in the Vpols. It should be noted that at the time of writing the Hpol design of RNO-G was not finalized and this analysis uses the performance parameters of a first iteration, so experiment related performance will change in the future.

Depending on the arrival direction, the radio signal will reach each antenna at a different time. For a full reconstruction of the neutrino properties, this timing offset has to be obtained from data. How to do this best is beyond the scope of this paper. We therefore calculate the correct time offset from the location of the simulated interaction vertex. Radio-based neutrino detectors have shown sub-nanosecond timing accuracy and the ability to reconstruct the signal direction to better than \SI{1}{^\circ}\cite{ARIANNA_air_shower_detection, ARA}, so correcting for time differences will likely not be the determining uncertainty.

For the purpose of demonstrating the IFT approach, we will only use the signals in the four Vpols making up the phased array and the two Hpols directly above them. These are located sufficiently close to each other that the radio signals reaching those antenna are very similar for most events. However, this is very dependent on the event geometry, so especially if the distance between the neutrino interaction vertex and the detector station is small, there can be differences in the signal, even between neighboring channels of the phased array. These manifest mostly as differences in the signal amplitude, while the shape of the spectrum tends to change less.

\subsection{Model Building}
\begin{figure}
    \centering
    \includegraphics[width=.95\textwidth]{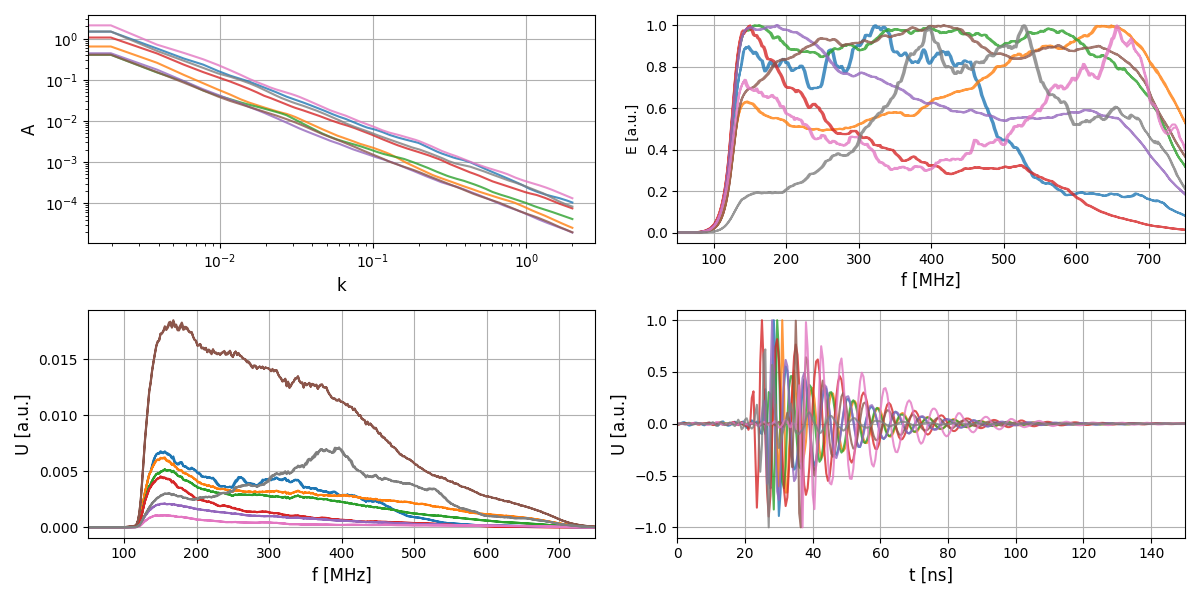}
    \caption{Samples drawn from the prior distribution of the IFT model. Top left: Amplitude power of the Fourier modes $k$ of the function describing the electric field spectrum. 
    Top right: Spectrum of the electric field. For better readability, all samples have been normalized to the same maximum value. Botton left: Spectrum of the channel voltage. Bottom right: Voltage trace in the time domain.}
    \label{fig:priors}
\end{figure}

Using IFT for the purpose of reconstructing short, non-repeating pulses, first a model of the expected pulse has to be developed. The approach to modelling the signal and the detector response is as follows: First, the amplitude and the complex phase factor of the electric field spectrum are modeled independently from each other as real and complex numbers in the frequency domain, and then multiplied together. Next, the response of the antenna and the amplifier, which are also modeled in the frequency domain, are applied to this to get the spectrum of the voltages measured by each channel. Finally, the result is transformed into the time domain where it can be compared to the data.

To have the electric field amplitude on a predictable scale, we normalize the waveforms of all used channels, so that their overall maximum value is equal to 1. As this normalization factor is known, it can be reapplied afterwards to obtain the correct electric field strength. Because the detector is only sensitive in a certain range of frequencies, we apply a 10th order Butterworth filter with an according passband (in this case \SI{132}{MHz} to \SI{700}{MHz}) to the channel waveform.

The electric field spectrum is split into its amplitude $E(f)$ and phase $\varphi (f)$:
\begin{equation}
    \mathcal{E} (f) = E (f) \cdot \exp(i \cdot \varphi(f))
\end{equation}
We model the amplitude $E(f)$ as a correlated, positive random process by means of a log-normal process, in which $s(f) = \log(E(f))$ obeys a Gaussian statistics. The correlation structure $S_{f,f'}:=\langle s(f) \, s(f') \rangle_{(s)}$ of the logarithmic amplitude is assumed to be statistically homogeneous in $f$, but the exact form of $S$ is unknown prior to observation. Following the prior model described in \cite{arras2020variable}, we express $S$ in terms of its eigen-spectrum $\tau(k)=|A(k)|^2$, with $k$ being the Fourier partner of $f$ and $A(k)$ an amplitude modulating function that turns standard white noise excitation variables $\xi'(k)$ into the Fourier space variables $s(k)=A(k)\,\xi'(k)$ that exhibit the correct statistics, 
$S_{k,k'}:=\langle s(k) \, s(k') \rangle_{(s)} = 2\pi\delta(k-k')\, \tau(k)$.
We place a prior on $\tau(k)$ such that it is preferably a power-law like spectrum $\propto k^{-\alpha}$, but allow for possible, differentiable deviations from this. 
We visualize this model by drawing random samples from the generative process and by plotting the results in Fig.~\ref{fig:priors}. The top left plot shows the spectral power modulation function $A(k)$ for the function describing the electric field amplitude.\footnote{Technically, this is a spectrum of a spectrum, and therefore $k$ is defined in the time domain. We forego this subtlety here and use the notation that is usually used for the power spectrum.} The slope $\alpha/2$ of this determines the smoothness of the electric field prior, with a steeper power spectrum (larger $\alpha$) resulting in a smoother prior. The resulting electric field spectra are shown in the top right of Fig.~\ref{fig:priors}. They are relatively smooth as a result of our choice of $\alpha$, but other than that no prior assumptions are made, except for the 10th order Butterworth filter that is applied at the edges of the detector's passband.
The small differences in the direction into which the radio signal reaching different channels was emitted can lead to small differences in the signal strength between channels. We want the model to be able to account for this. In case the signal-to-noise ratio is relatively small, we do not expect to be sensitive to such differences anyway, and would rather avoid adding additional free parameters. Anticipating the variation between channels is rather difficult as this is very dependent on the event geometry, so we opt for a simple measure here:
If the signal-to-noise ratio (SNR)\footnote{We define the signal-to-noise ratio as half the peak-to-peak amplitude, divided by the root-mean-square of the noise.} of any of the measured waveforms is above 10, we multiply the spectrum amplitude for each channel with an additional parameter $a$. This parameter's Gaussian prior distribution is centered around 1 with a standard deviation of $\sigma_a=0.1$, which means that we assume the difference between channels to be relatively small, unless the data shows otherwise. If none of the channels reaches an SNR of 10, this parameter is left out, which forces the electric field at each channel to be the same. Whether or not to include this parameter is ultimately a judgement call. In general, it is better to avoid unnecessary free parameters in the model, on the other hand measuring these differences between channels provides additional information that could be used for the neutrino reconstruction and may be necessary for the IFT model to be able to fit the data. A simple SNR cut to decide this is rather crude and somewhat arbitrary, but it is good enough for our purposes here and may easily be substituted for a more advanced decision process, should it become necessary.

The electric field consists of a single, very short pulse. Modelling the phase as constant over the entire spectrum would result in a pulse at $t=0$ that rolls over into the end of the waveform. Since a shift in the time domain corresponds to a slope of the phase in the frequency domain, we can shift the model pulse to the correct position in the waveform, by modelling the phase $\varphi(f)$ as a linear function
\begin{equation}
    \varphi(f) = \varphi_0 + m f
\end{equation}
where $\varphi_0$ and $m$ are Gaussian variables. By choosing the mean $\bar{m}$ and standard deviation $\sigma_m$ of the slope $m$, we can encode the position of the pulse in the recorded waveform as well as the uncertainty on it. This is useful, since the exact position of the pulse maximum can depend on the amplitude of the spectrum as well as the phase, making it tricky to determine the correct slope that should be used for the phase model ahead of time. Group delay in the antenna or amplifier response can also introduce small time offsets. With this fitting procedure, small timing errors of a few nanoseconds can be adjusted for.

Thus, the full prior distribution for $\mathcal{E}(f)$ is given as
\begin{eqnarray}
	\mathcal{P}\left(\mathcal{E}(f)\right) &= \int ... \int& \delta\left(\mathcal{E}(f) - e^{s(f) + i \left(\varphi_0 + m f\right)}\right) \mathcal{N}\left(s|0, S(\tau)\right) \ \mathcal{P}(\tau) \nonumber \\ &&\mathcal{N}\left({\varphi}_0|\bar{\varphi}_0, \sigma_0^2\right) \ \mathcal{N}\left(m|\bar{m}, \sigma_m^2\right) \mathrm{d}s \mathrm{d}\tau \mathrm{d}\varphi_0 \mathrm{d}m  .
\end{eqnarray}

The spectra of the $\theta$  and $\varphi$ components of the electric field are identical to each other in shape, so the polarization can be described by a single parameter, called \emph{polarization angle} defined as $\tan(\varphi_{pol}) = \frac{|\mathcal{E}_\phi|}{|\mathcal{E}_\theta|}$. This allows us to split the electric field spectrum into its $\vec{e}_{\theta}$ and $\vec{e}_{\phi}$ components:
\begin{equation}
    \vec{\mathcal{E}}(f) = \mathcal{E}(f) \cdot \cos(\phi_\mathrm{pol}) \cdot \vec{e}_\theta + \mathcal{E}(f) \cdot \sin(\phi_\mathrm{pol}) \cdot \vec{e}_\phi
\end{equation}
We want avoid to restrict the polarization a priori, so we model its prior probability with a Gaussian, but set the standard deviation $\sigma_\phi$ so large that it has negligible influence on the posterior distribution. This would in principle allow us to later on incorporate prior knowledge by decreasing $\sigma_\phi$ and setting the mean accordingly, if additional information is available, for example from the event geometry.

The antenna response to the radio signal is expressed through the vector effective length 
\begin{equation}
    \vec{\mathcal{H}}_i(f, \theta, \phi)=\mathcal{H}_i^{\theta}(f,\theta, \phi)\cdot\vec{e}_\theta+\mathcal{H}_i^\theta(f, \theta, \phi)\cdot\vec{e}_\phi,
\end{equation}
with $f$ being the frequency, $\theta$ and $\phi$ the zenith and azimuth angles of the incoming radio signal and $\vec{e}_\theta$ and $\vec{e}_\phi$ their respective base vectors. The voltage over the antenna output is then given by
\begin{equation}
    \mathcal{V}_i^\mathrm{ant}(f)=\vec{\mathcal{E}}(f) \cdot \vec{\mathcal{H}}_i(f, \theta, \phi)=\mathcal{E}(f) \cos(\phi_\mathrm{pol}) \mathcal{H}_i^{\theta}(f, \theta, \phi) + \mathcal{E}(f) \sin(\phi_\mathrm{pol}) \mathcal{H}_i^{\phi}(f, \theta, \phi)
\end{equation}
The complex amplifier response $A_i(f)$ can also be expressed by a multiplication in the frequency domain, yielding the voltage as it is recorded by the digitizer:
\begin{equation}
    \mathcal{V}_i(f)= A_i(f) \cdot \mathcal{V}_i^\mathrm{ant}(f)=A_i(f)\cdot \mathcal{E}(f)\cdot [\cos(\phi_\mathrm{pol}) \cdot \mathcal{H}_i^\theta(f, \theta, \phi)+ \sin(\phi_\mathrm{pol})\cdot \mathcal{H}_i^\phi(f, \theta, \phi)]
\end{equation}
The product of the amplifier and the antenna responses has been normalized, so that their overall maximum value (over all channels) is 1. As with the normalization of the voltage waveforms this can be reversed afterwards in order to get the correct values for the electric fields.

As these calculations are just a series of multiplications of each element of $\mathcal{E}(f)$ with a scalar, they can be expressed as a diagonal operator $\mathcal{D}$ that is applied to the electric field spectrum. The result is shown in the bottom left of Fig.~\ref{fig:priors}. The prior samples for the voltage spectra look much more similar to each other than the electric field spectrum samples. This is because the antenna and amplifier characteristics have a large influence on the shape of the spectrum.
Finally, the voltage $\mathcal{V}(f)$ is transformed into the time domain by applying the Fourier transformation operator $\mathcal{F}$ (Fig.~\ref{fig:priors}, bottom right). The time shifts between the samples is due to the allowed uncertainty on the slope of the phase function.

This allows us to calculate the probability $\mathcal{P}(U|E)$ that a given electric field model yields the voltage measured by the detector. As any discrepancies between the voltage expected from the electric field model $\mathcal{F} \mathcal{DE}(f)$ and the actual data must be due to the noise $N$, this means:
\begin{equation}
    \mathcal{P}(U|E) = \mathcal{P}(N=U-\mathcal{F}\mathcal{DE})
\end{equation}
If we have a probability for the noise distribution (which will be presented later) this allows us to calculate the posterior probability for a given electric field:
\begin{equation}
\mathcal{P}(E|U) \propto \mathcal{P}(E) \cdot \mathcal{P}(N=U-\mathcal{F}\mathcal{DE})
\label{eq:efield_posterior}
\end{equation}

\subsection{Use case: Air shower detection}
The radio signals from cosmic ray-induced particle showers in the atmosphere are produced by similar mechanisms to those from neutrino-induced showers in ice and share many features. Therefore, IFT can be used here as well to reconstruct the electric field. Additionally, since radio signals from air showers are better understood, they are offer a useful cross-check to test reconstruction methods for neutrino detectors.

While RNO-G has three upward-pointing LPDAs for air shower detection, we will use a different neutrino detector as a test case. The ARIANNA experiment \cite{ARIANNAdesign} consists of 9 autonomous detector stations located on the Ross Ice Shelf in Antarctica and 2 more at the South Pole. Each station has 4 or 8 (depending on the station type) LPDAs, similar to those used for RNO-G, deployed at a depth of about \SI{2}{m} below the ice surface. Two each of the antennas form a pair, which means that they are oriented parallel to each other, so that they measure the same component of the electric field. While most of the stations have their antennas pointed downwards, to detect radio signals coming from within the ice, they can be turned upwards as well, which turns the station into a cosmic ray detector. This was already used in the past to detect several cosmic ray events \cite{ARIANNA_air_shower_detection}.

In addition to cosmic ray detection at the Ross Ice Shelf, we want to simulate another scenario as well: The reconstruction of inclined air showers at the site of the Pierre Auger Observatory in Argentina. The observatory is currently undergoing an upgrade, in which radio antennas will be installed on all water Cherenkov detector tanks \cite{auger_prime}. Because of the large distance between the antennas, the radio array will be mainly focused on inclined showers, whose radio footprint on the ground is large enough to cover multiple stations. With antennas only sensitive to the horizontal components of the radio signal, reconstructing the electric field can be challenging for the reasons explained in Sec.~\ref{sec:classic_reconstruction}. In the arctic or antarctic, there is a trivial solution for this problem: The geomagnetic field is almost vertical, which means that the $\vec{e}_\varphi$ component is dominant, so the $\vec{e}_\theta$ component can just be ignored when estimating the energy fluence of the electric field. This is not necessarily the case at the Auger site, where the magnetic field is only inclined around \SI{30}{^\circ} and weaker. Thus, we want to show how this problem can be addressed via IFT.

Since all the assumptions made about the signal from in-ice showers also hold for air showers, only a few modifications to account for the different detector design are necessary: The antenna and amplifier response are simply replaced with the ones for the ARIANNA detector. They are sensitive at different frequencies, so the passband of the Butterworth filter that is applied is changed to \SI{100}{MHz} to \SI{500}{MHz}. Additionally, part of the radio signal will be reflected at the air-ice interface, but since the fraction of the signal that is reflected can be calculated separately for the $\vec{e}_\theta$ and $\vec{e}_\phi$ component, and only depends on the incident angle otherwise, this is done by simply modifying the antenna response. The radio footprint of an air shower at ground level is so large compared to the distance between the antennas that the scaling factor $a$ between channels can be ignored.

\subsection{Noise models}
\label{sec:noise}
Modelling the noise for a detector at a given location is challenging, as there are typically many different sources: Foremost human activity, but also emission from the Galactic center or other celestial objects as well as thermal emission from the surrounding ice or the detector itself can all contribute different forms of noise.

For the use case of air shower detection with ARIANNA, one can simply use real data: Each detector station is regularly triggered on a timer. Since actual signal events are very rare, it can be assumed that these forced trigger events contain only noise, and their voltages can simply be added to the simulated waveforms in order to obtain simulated noisy data.

The noise distribution obtained from the ARIANNA forced trigger events is nearly Gaussian. This is likely due to the fact that most of the noise is thermal noise produced in the amplifiers, rather than impulsive signals picked up by the antennas from the surroundings.

This is not necessarily going to be the case for RNO-G, whose amplifiers are targeting a lower noise level. Its location in the arctic also means that the galactic center, which is a major source of radio emission in the sky, is not within the field of view. We therefore assume that most of the noise comes from thermal radio emission which is picked up by the antennas. We model this noise in the frequency domain, with a Rayleigh distribution on the amplitude of each frequency bin and a uniform distribution over $0 < \varphi \leq 2\pi$ of the phase and the polarization angle. In principle, antennas at different depths could detect different noise, as radio signals from the surface get attenuated and the temperature of the surrounding ice changes. For simplicity, we here assume the noise level to be the same for all channels.

In IFT, the easiest way to model noise is to assume a Gaussian distribution around 0, with a standard deviation $\sigma_N$, which is set for each channel individually by calculating the root mean square of the voltages outside the signal region. Thus we can substitute the noise term in Eq.~\ref{eq:efield_posterior} with a Gaussian distribution $\mathcal{P}(N) = \mathcal{N}(N|0, \sigma_N)$.

This means we are technically using a slightly wrong noise model for a typical radio neutrino detector. However, it turns out that this does not hinder a precise electric field reconstruction and serves to demonstrate the resilience of our reconstruction to small errors in the noise model. In principle, our measurement model can easily accommodate more complex noise distributions, should the need arise.


\section{Performance of the electric field reconstruction}

With the posterior probability in Eq.~\ref{eq:efield_posterior}, we can reconstruct the electric field by finding its maximum. We do this using the software package \texttt{nifty} \cite{asclnifty5}, which provides the necessary tools to formulate the described IFT models and perform the variational inference.
In this chapter, we create realistic datasets from simulated neutrino and air shower signals and evaluate the 
ability of \texttt{nifty} to recover the original signal using our signal and data model.

\subsection{Set-up of neutrino simulations}
\begin{figure}
    \centering
    \includegraphics[width=.9\textwidth]{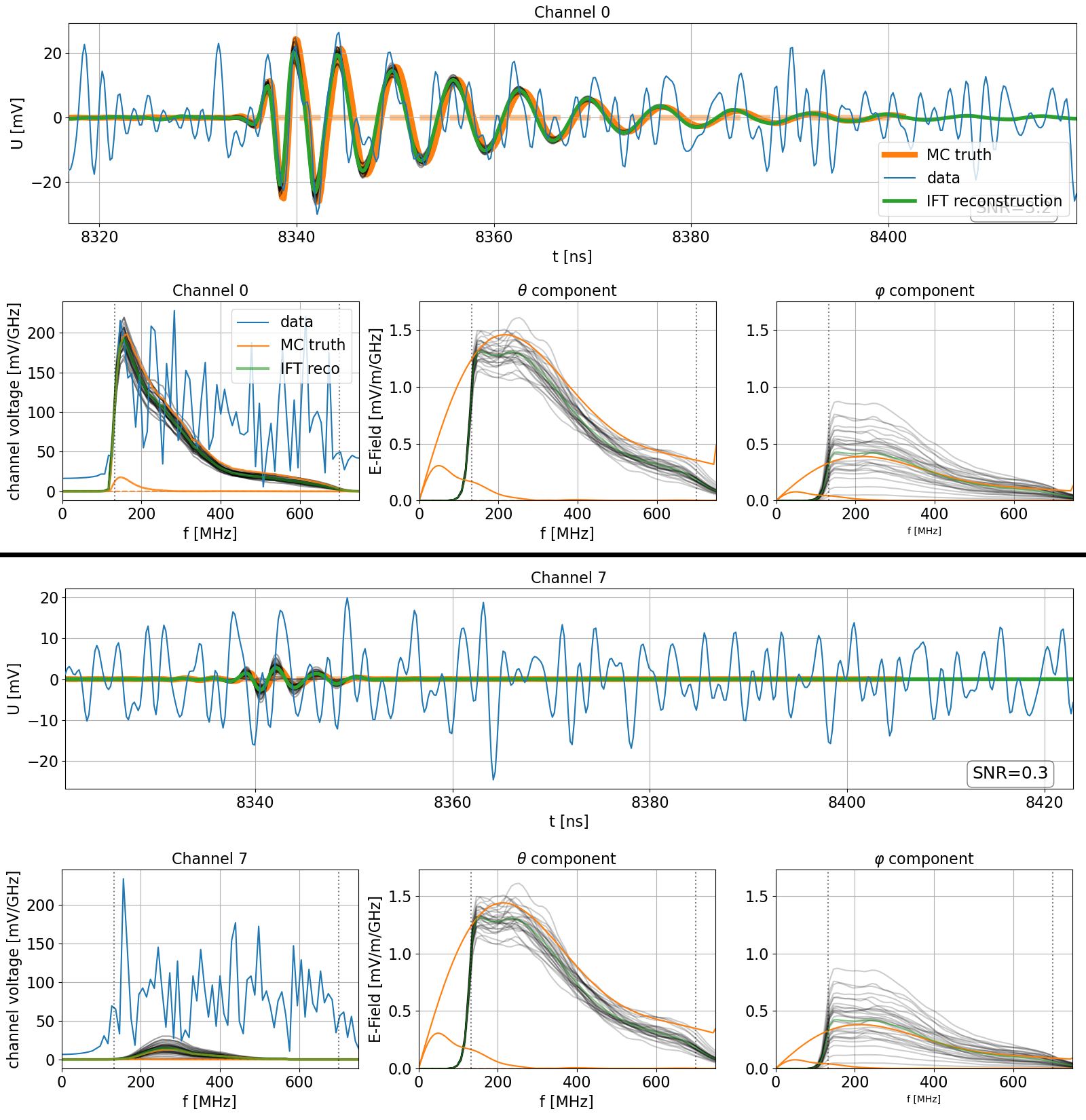}
    \caption{A result of an electric field reconstruction, shown for one of the Vpol (top) and one of the Hpol (bottom) antennas. 1st and 3rd row: Waveforms of the noisy data (blue), the noiseless signal (orange) and the reconstructed voltages (green). 2nd and 4th row: Spectrum of the channel voltage (left) and of the $\vec{e}_\theta$ (center) and $\vec{e}_\phi$ (right) component of the electric field. The thin black lines show posterior samples, drawn to visualize and quantify the uncertainties of the reconstruction. The 2nd orange line in the spectrum plots is a 2nd radio pulse from another raytracing solution, potentially visible in the waveform.}
    \label{fig:reco_example}
\end{figure}

We use the NuRadioMC and NuRadioReco software packages \cite{NuRadioMC, NuRadioReco}, to simulate the radio signals from neutrino-induced particle showers detected by an RNO-G station. 
In the first step, neutrino interactions are randomly generated in a cylinder of radius \SI{5}{km} around the station up to a depth of \SI{3}{km}, corresponding roughly to the thickness of the ice sheet at the RNO-G site. The neutrino energy is chosen randomly between \SI{5.e16}{eV} and \SI{1.e20}{eV}, following the combined spectrum of a GZK flux model \cite{vanVliet:2019nse} and the extension of the astrophysical neutrino spectrum measured by IceCube \cite{icecube_flux}. Neutrino flavors are assigned assuming a 1:1:1 mixing between the three flavors. In the second step, the radio emission is simulated using the ARZ model \cite{Alvarez_Mu_iz_2010} and the signal is propagated through the ice. If the radio signal reaches the detector, the electric field is folded with the antenna and the amplifier response, which yields the voltage measured by each channel. We simulate a simple threshold trigger, that triggers if the maximum voltage in the lowest channel of the phased array is above \SI{20}{mV}.

If the detector is triggered, we add noise with a root mean square of \SI{10}{mV}, as described in Sec.~\ref{sec:noise}. Then the signal is upsampled to a sampling rate of \SI{5}{GHz} and a 10th order butterworth filter with a passband of \SIrange{132}{700}{MHz}, corresponding to the band where antenna and amplifier are sensitive, is applied.

Finally, we perform the electric field reconstruction on the resulting waveforms, using the four channels of the phased array and the two horizontal polarization antennas on the power string.

\subsection{Test of the electric field reconstruction}
The result of one of these reconstructions is shown in Fig.~\ref{fig:reco_example}. We show only one of the Vpol and one of the Hpol channels, as the reconstruction results and the recorded waveforms are practically identical to the other channels, except for different noise. The reconstructed waveform matches the original signal very closely, even towards the end of the pulse, where it drops below the noise level. This is possible because much of the pulse shape is dictated by the group delay of the antenna and the amplifier, which is well known. We can estimate lower bounds on the uncertainties by drawing samples from the approximate posterior distribution provided by MGVI. The samples are shown in Fig.~\ref{fig:reco_example} as thin black lines, to visualize their variance. The uncertainty on any signal property, like for example the 
polarization, can be characterized thereby. For this, one calculates this quantity for each sample individually and investigates its resulting distribution, for example by providing  its standard deviation as calculated from the samples.

\begin{figure}
    \centering
    \includegraphics[width=.9\textwidth]{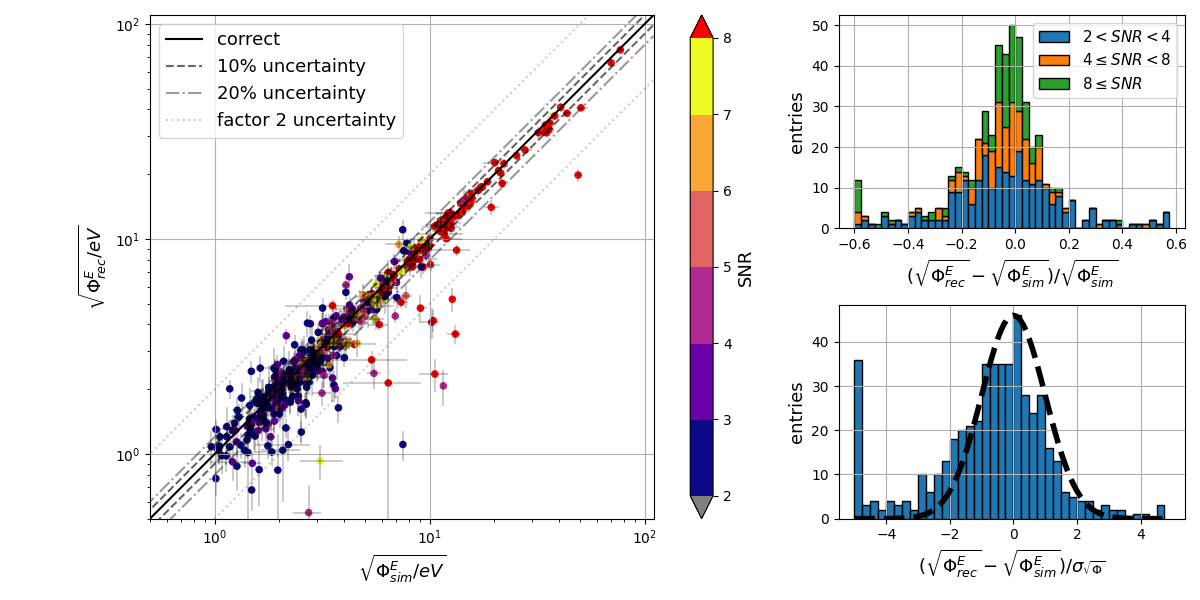}
    \caption{Reconstructed energy fluence of the radio signal from simulated neutrino-induced particle showers detected with RNO-G. Left: Scatter plot of the reconstructed vs.\ the true square root of the energy fluence of the radio signal. Errorbars on the y-axis show the uncertainty estimated by the IFT algorithm, errorbars on the x-axis cover the range of values for the electric fields at all channels used in the reconstruction, with the data point being at the mean. The colors show the maximum signal-to-noise ratio of all channels used in the reconstruction. Top right: Stacked histogram of the relative uncertainty on $\sqrt{\Phi^E}$ for different signal-to-noise ratios. Bottom right: Uncertainties on the reconstructed $\sqrt{\Phi^W}$ as a multiple of the estimated uncertainty. For comparison, a Gaussian function with $\sigma=1$ as drawn as well. The left- and rightmost bins in each histogram are overflow bins.}
    \label{fig:fig:energy_fluence_rno}
\end{figure}
We want to quantify the performance of the electric field reconstruction by looking at properties of the radio signal that are important for the reconstruction of the neutrino that produced the particle shower the radio signal came from. How well these are reconstructed will depend mainly on the signal-to-noise ratio, so we will show results divided into separate groups by the maximum SNR of any of the channels used in the reconstruction. The SNR is calculated by taking half the maximum peak-to-peak amplitude of the signal \emph{before the noise was added} and dividing it by the root mean square of the noise that is added.
The reason we use the waveform without noise to calculate the amplitude is that otherwise, it would be influenced by the noise as well as the underlying signal. If the signal is not much bigger than the noise level, its value is mostly influenced by random fluctuations in the noise. So, even though it is harder to calculate for real data, this definition makes it easier to compare low-SNR events when judging the performance of the electric field reconstruction. The trigger threshold and simulated noise level have been chosen so that the lowest SNR are around 2, which is also the expected trigger threshold for RNO-G.

The first quantity to check is the energy fluence $\Phi^E$ of the radio signal (see Eq.~\ref{eq:energy_fluence}), which is needed to reconstruct the shower energy, from which the neutrino energy can be estimated. We look at the reconstruction quality of $\sqrt{\Phi^E}$, which is proportional to the shower energy, so the relative uncertainty on $\sqrt{\Phi^E}$ propagates directly to the uncertainty on the shower energy.

The result is shown in Fig.~\ref{fig:fig:energy_fluence_rno}. On the left, it shows a scatter plot of the reconstructed vs.\  actual $\sqrt{\Phi^E}$. Except for a few outliers, the reconstruction is very close to the actual value, with a resolution on $\sqrt{\Phi^E}$ of around 20\% even for events with $\mathrm{SNR} < 4$ (see Tab.~\ref{tab:resolution_rno}). Because each antenna views the shower at a slightly different angle, the electric field at each antenna can have a different energy fluence, especially if the shower is relatively close. The horizontal errorbars extend from the minimum to the maximum energy fluence of all six channels used for the reconstruction, with the data point being at the mean. This also means that the channels can have different signal-to-noise ratios, so the colors show the maximum SNR of all channels used in the reconstruction.
For most events, the uncertainty on the energy fluence of the $\vec{e}_\phi$ component is larger than on the $\vec{e}_\theta$ component. This is because there are only two Hpol antennas, compared to the four Vpols, which also are less sensitive. Together with the fact that the signals tend to be polarized more vertically, this means that often there is only a very small signal in the waveforms recorded by the Hpol channels. In that case, only an upper limit on the $\vec{e}_\phi$ component of the electric field can be given. This is also the reason why the energy fluence tends to be more often under- than overestimated, as shown in Tab.~\ref{tab:resolution_rno}.

The IFT reconstruction algorithm also provides an estimate of the uncertainty $\sigma_{\sqrt{\Phi}}$ on $\sqrt{\Phi^E}$, shown as vertical errorbars. The bottom right of Fig.~\ref{fig:fig:energy_fluence_rno} shows the difference between reconstructed and actual $\sqrt{\Phi^E}$ divided by $\sigma_{\sqrt{\Phi}}$. It follows a Gaussian distribution with $\sigma=1$, as would be expected if the estimated uncertainties are accurate.

There are some outliers, even at large signal-to-noise ratios, where the energy fluences are greatly underestimated. These can occur if either the data does not match the IFT model (which can happen for a few reasons) or if the reconstruction converges on a local minimum, usually because of a wrong phase. Fortunately, these errors are usually easy to spot on an individual event level and can be cut before an analysis. They are discussed in more detail in App.~\ref{sec:failed_recos}.

\begin{figure}
    \centering
    \includegraphics[width=.95\textwidth]{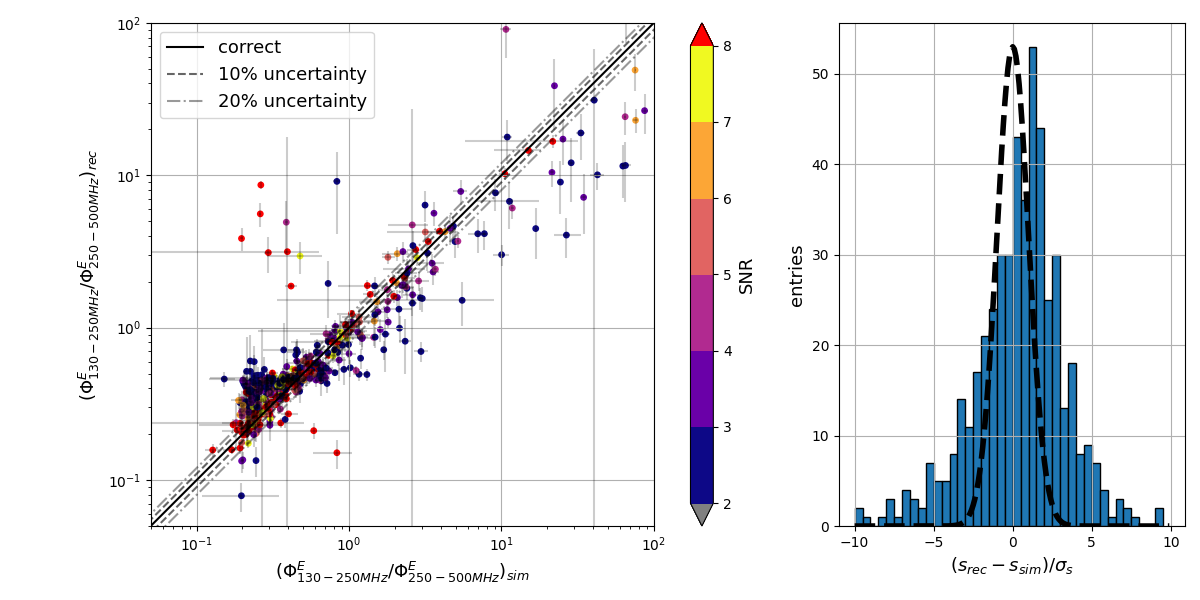}
    \caption{Reconstruction of the spectral shape parameter $s$ for radio signals detected with RNO-G. Left: Reconstructed vs.\ true simulated $s$. Horizontal errorbars show the range of values for all channels used for the reconstruction, vertical errorbars show uncertainty estimates by the reconstruction algorithm. Right: Uncertainties on the reconstructed $s$ divided by the uncertainties estimated by the reconstruction algorithm.}
    \label{fig:slope_scatter_rno}
\end{figure}
To reconstruct the shower energy and the neutrino direction, it is also important to know the shape of the electric field spectrum, as this contains information about the angle under which the detected radio signal was emitted. The frequency at which the electric field spectrum reaches its maximum varies a lot, depending mostly on the viewing angle and shower type (see Fig.~\ref{fig:signal_examples}) and may even be outside of the detector bandwidth, but for signals that are off-cone by more than \SI{1}{^\circ} it tends to be somewhere around \SI{250}{MHz}. Therefore we compare the energy fluence in the frequency bands at \SIrange{130}{250}{MHz} and at \SIrange{250}{500}{MHz} to give an estimate on how quickly the spectrum drops off after the maximum (if it drops at all). The reason for not using the whole bandwidth up to \SI{700}{MHz} is that the antenna response depends on the incoming direction of the radio signal, and for some directions the sensitivity starts dropping at about \SI{500}{MHz} (see Fig.~\ref{fig:rno_station}). From these fluences we calculate a shape parameter $s=\Phi^E_{130-200\mathrm{MHz}}/\Phi^E_{200-500\mathrm{MHz}}$, which we compare to the one from the simulated electric fields. This estimator for the viewing angle has the advantage that it is well-defined for any electric field spectrum. If we, for example, had used the position of the maximum of the spectrum, this would we be ill-defined for events where it is outside the RNO-G band or if there are multiple peaks.

As Fig.~\ref{fig:slope_scatter_rno} shows, the IFT method is able to reconstruct the shape rather well for events with $\mathrm{SNR} > 3$, with the exception of some outliers. Uncertainties tend to be larger than the lower bound estimated by the IFT algorithm. One may also notice that many of the events with a low signal-to-noise ratio are clustered in the lower left corner, with roughly the same reconstructed values for $s$. These are events that are viewed relatively close to the Cherenkov angle, for which the IFT reconstruction returns a flat frequency spectrum, as there is not enough information for a more detailed spectrum reconstruction. This result is useful though, since the falling spectrum of events further off the Cherenkov angle is still reconstructed well, so some information about the viewing angle is maintained.

\begin{figure}
    \centering
    \includegraphics[width=.9\textwidth]{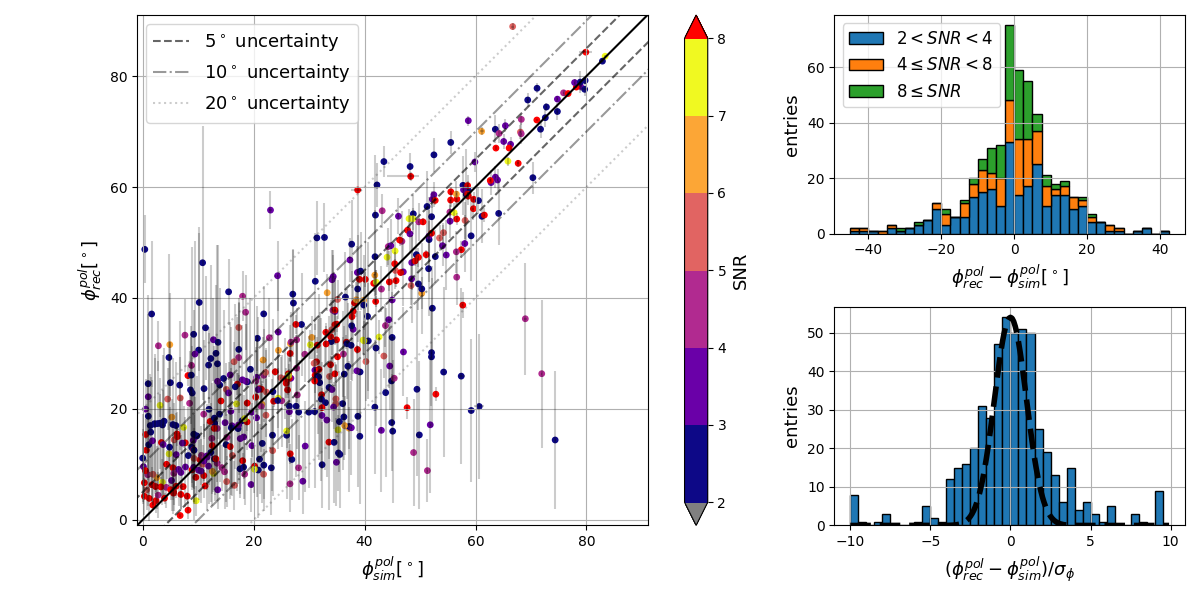}
    \caption{Reconstructed electric field polarization of the simulated radio signals detected with RNO-G. Left: reconstructed vs.\ true polarization angles. Errorbars represent the uncertainties estimated by the IFT reconstruction algorithm. Colors signify the maximum signal-to-noise ratio out of all channels used in the reconstruction. Top right: Stacked histogram of the uncertainties on the reconstructed polarization angles for different signal-to-noise ratios. Bottom right: Uncertainties on the reconstructed polarization angle divided by the estimated uncertainties. For comparison, a Gaussian function with $\sigma=1$ has been drawn as well.}
    \label{fig:polarization_scatter_rno}
\end{figure}

The third parameter to look at is the polarization of the radio signal. The electric field vector points towards the shower axis, which helps to constrain the direction the neutrino came from, which makes the polarization an important quantity for the reconstruction of the arrival direction. We quantify the polarization by defining a polarization angle 
\begin{equation}
    \phi^\mathrm{pol} = \tan^{-1}( \sqrt{\Phi^E_{\phi}} / \sqrt{\Phi^E_{\theta}})
\end{equation}
where $\Phi^E_{\phi}$ and $\Phi^E_{\theta}$ are the energy fluence of the $\vec{e}_\phi$ and $\vec{e}_\theta$ components of the radio signal. This implies that the polarization angle lies in the range $0^\circ \leq \phi^\mathrm{pol} \leq 90^\circ$ by definition, where $\phi^\mathrm{pol}=0^\circ$ means that the signal is polarized in $\vec{e}_\theta$ direction.

The results are shown in Fig.~\ref{fig:polarization_scatter_rno} and Tab.~\ref{tab:resolution_rno}. Even at signal-to-noise ratios below 4, the polarization can be reconstructed with an uncertainty of around \SI{13}{^\circ}, which improves to better than \SI{5}{^\circ} for $\mathrm{SNR} > 8$. The resolution of the polarization measurement is mostly limited by the lower sensitivity and smaller number of the Hpol antenna, as even a large SNR in the Vpol channels does not necessarily mean that there is also a large SNR in the Hpol channels. This is also the reason why the polarization reconstruction performs better for larger $\phi^{pol}$: If the polarization angle is small, there is no detectable or only a very weak signal in the Hpol antenna waveform, so in practice, only an upper bound on $\Phi^E_{\phi}$ can be estimated.

The errors on the reconstructed polarizations match the IFT estimates for the lower bound on the uncertainty rather well. It is worth mentioning that there is a potential systematic error on the polarization due to the detector layout: While the differences in signal polarization between channels are small enough to be insignificant, the overall signal strengths may vary. Because the $\vec{e}_\theta$ and $\vec{e}_\phi$ components of the electric field are measured by different antennas, this variation leads to wrong polarization reconstruction. This is a limitation of the detector, which is ignored in the IFT model, so it cannot be included in the uncertainty estimate.

\begin{table}[]
    \centering
    \begin{tabular}{|c|c|c|c|}
         \hline
         & $2 \leq \mathrm{SNR} < 4$ & $4 \leq \mathrm{SNR} <8$ & $8 < \mathrm{SNR}$  \\
         \hline
         $\Delta \Phi^E$ & && \\
         68\% quantile & [-0.37, 0.34] & [-0.30, 0.11] & [-0.28, 0.05] \\
         median & -0.06 & -0.07 & -0.06  \\
         \hline
         $\Delta \sqrt{\Phi^E}$ &&&\\
         68\% quantile & [-0.20, 0.16] & [-0.16, 0.05] & [-0.15, 0.03] \\
         median & -0.03 & -0.04 & -0.02 \\
         \hline
         $\Delta \phi_\mathrm{pol}$ &&& \\
         68\% quantile & [\SI{-12.4}{^\circ}, \SI{13.8}{^\circ}] & [\SI{-9.5}{^\circ}, \SI{7.8}{^\circ}] & [\SI{-4.9}{^\circ}, \SI{4.2}{^\circ}] \\
         median & \SI{0.3}{^\circ} & \SI{0.7}{^\circ}& \SI{0.0}{^\circ} \\
         \hline
    \end{tabular}
    \caption{Uncertainties on the reconstructed signal energy fluence $\Phi^E$, its square root and the polarization for radio signals by an RNO-G station. For each property, the 68\% quantile and the median of the distribution of reconstruction errors are given.}
    \label{tab:resolution_rno}
\end{table}

\subsection{Reconstruction of pulses from air showers}
\begin{figure}
    \centering
    \includegraphics[width=.9\textwidth]{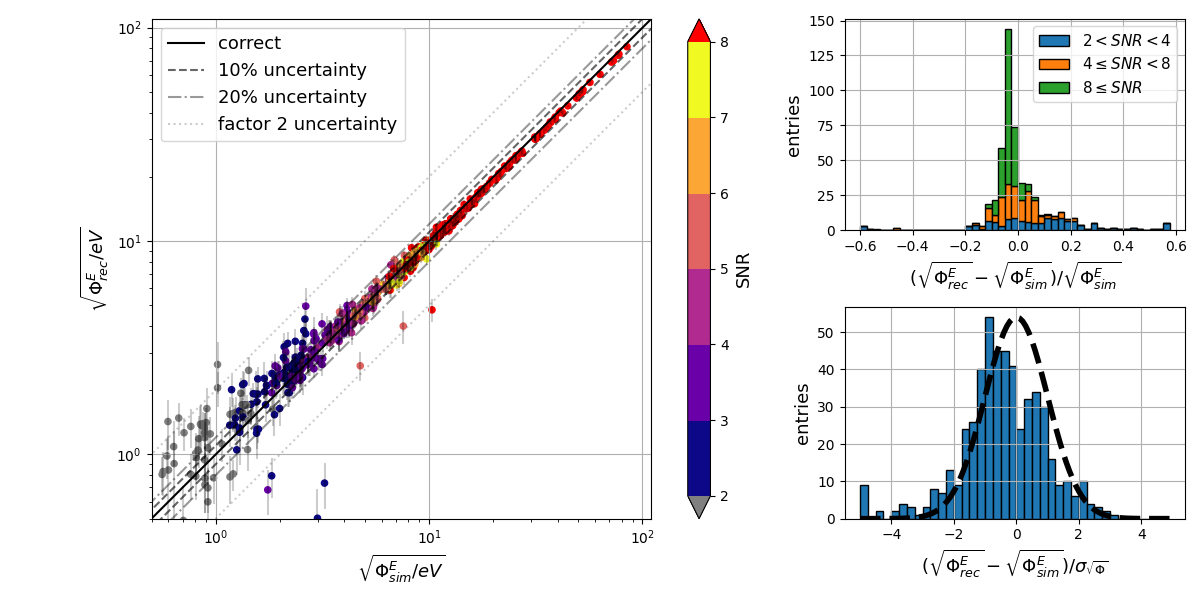}
    \caption{Reconstruction of the energy fluence of radio signals from air showers detected with an ARIANNA-like detector at Moore's Bay. Left: Reconstructed vs.\ actual $\sqrt{\Phi^E}$ of the radio signal. The vertical error bars show the uncertainties estimated by the IFT reconstruction algorith. Top right: Error on the reconstructed $\sqrt{\Phi^E}$ of the radio signal for different signal-to-noise ratios. Bottom right: Errors on the reconstructed $\sqrt{\Phi^E}$ divided by their estimated uncertainties.}
    \label{fig:energy_fluence_mooresbay}
\end{figure}
To test the IFT reconstruction of radio pulses from air showers, we use two sets, each containing 70 simulated air showers and their radio emission, generated with the CoREAS software \cite{Coreas}: One for a detector on the Ross-Ice-Shelf in Antarctica and one at the site of the Pierre Auger Observatory in Argentina. The main differences between these two locations are the different geomagnetic field and the different altitude, which is sea level at the Ross Ice Shelf and \SI{1560}{m} a.s.l. at the Auger site. The cosmic ray energies are randomly distributed between \SI{5.e17}{eV} and \SI{1.e20}{eV} with a spectral index of $-2$. The cosmic ray directions are distributed isotropically, with the air showers at the Auger site restricted to zenith angles $\theta > 50^\circ$, as we want to test the suitability of the method for inclined showers specifically. For each shower, the electric field is simulated at a set of positions arranged around the shower core in a star-like pattern (see e.g. \cite{Buitink:2014eqa}). 
For each simulated shower, a set of 20 positions around the shower core is chosen at random and the station positions closest to each position are selected, which results in a realistic distribution of shower axis positions relative to the station.
The electric field is folded with the response of each antenna and amplifier to get the waveform recorded by each channel. As we do not have the detector response of the Auger radio component, we use the ARIANNA detector for both locations. A time shift is applied to the waveforms to account for the difference in travel time between antennas.
We simulate a trigger by requiring that the signal in at least two channels passes both a minimum and a maximum threshold of \SI{-20}{mV} and \SI{20}{mV}, respectively, within \SI{5}{ns} of each other. This is similar to the trigger used in the ARIANNA experiment \cite{ARIANNA_air_shower_detection}, except that the threshold corresponds to roughly $2\sigma$, compared to the $4\sigma$ trigger that is actually used. We chose a lower threshold, because we are triggering on the noiseless waveforms, so pure noise triggers are not an issue, and we are especially interested in the performance of the IFT method at low signal-to-noise ratios.

The waveforms for each channel are cut to a length of 256 samples, to match the length of the noise samples measured by the ARIANNA detector and a randomly selected noise recording is added to each channel. We do not have any recorded noise from the Auger site, so we use recordings from Moore's Bay for both detector locations. The waveforms are upsampled to a sampling rate of \SI{5}{GHz} and a 10th order butterworth filter with a passband of \SI{100}{MHz}-\SI{500}{MHz} as applied. Then the electric field reconstruction is performed. We use the settings as for the neutrino case, except for the slope of the phase. This is because the ARIANNA antennas and amplifiers introduce a different delay to the radio signal, which has to be corrected for.

\begin{figure}
    \centering
    \includegraphics[width=.9\textwidth]{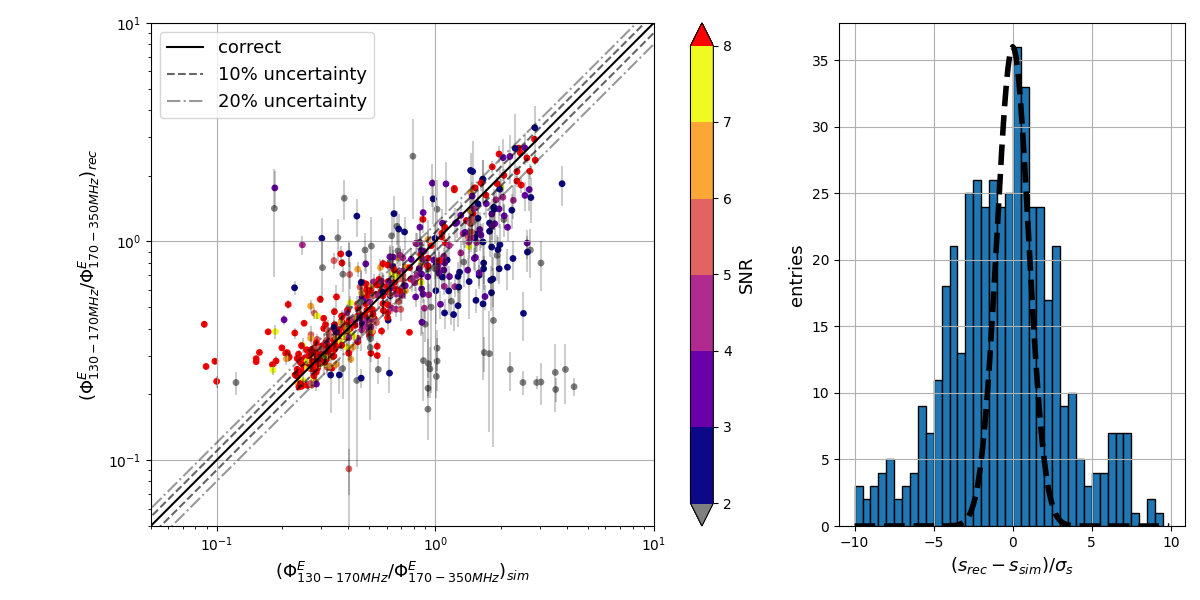}
    \caption{Reconstruction of the spectral shape parameter $s$ for radio signals from air showers detected with an ARIANNA station at Moore's Bay. Left: Reconstructed vs.\ actual shape parameter. Error bars show the uncertainties estimated by the IFT algorithm. Right: Errors on the reconstructed shape parameter divided by its estimated uncertainties.}
    \label{fig:slope_mooresbay}
\end{figure}

\begin{figure}
    \centering
    \includegraphics[width=.9\textwidth]{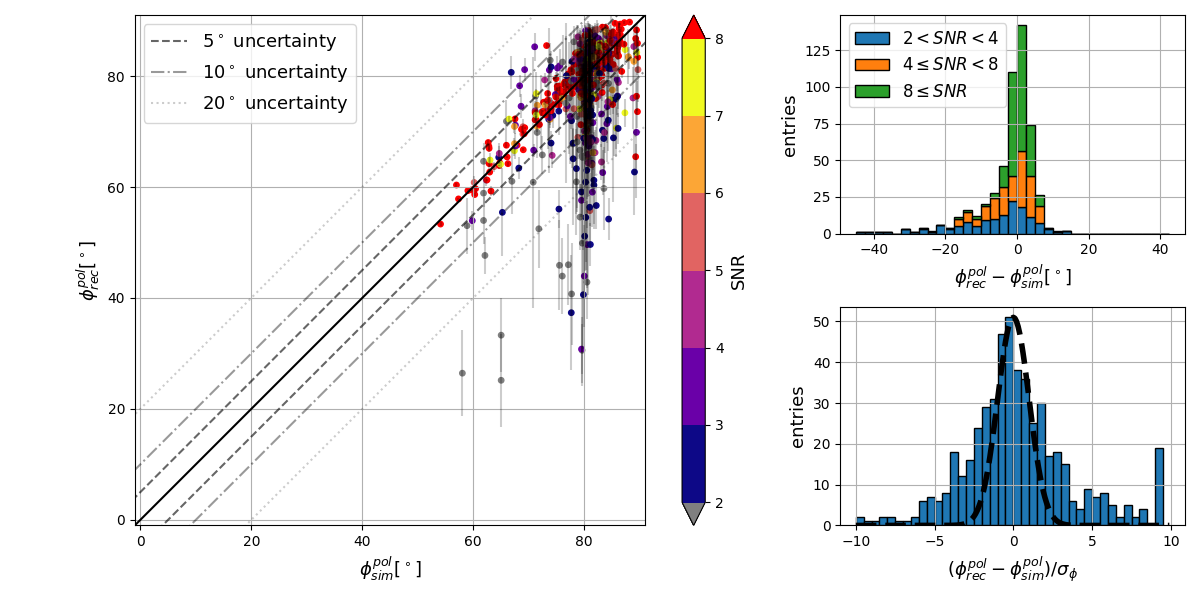}
    \caption{Reconstruction of the polarization for radio signals from air showers detected with an ARIANNA-like detector at Moore's Bay. Left: Reconstructed vs.\ actual polarization angle. Error bars show the uncertainties estimated by the IFT algorithm. Top right: Stacked histogram of the error on the reconstructed polarization angle for different signal-to-noise ratios. Bottom right: Errors on the reconstructed polarization angles divided by the estimated uncertainties.}
    \label{fig:polarization_moorebay}
\end{figure}

\begin{figure}
    \centering
    \includegraphics[width=0.9\textwidth]{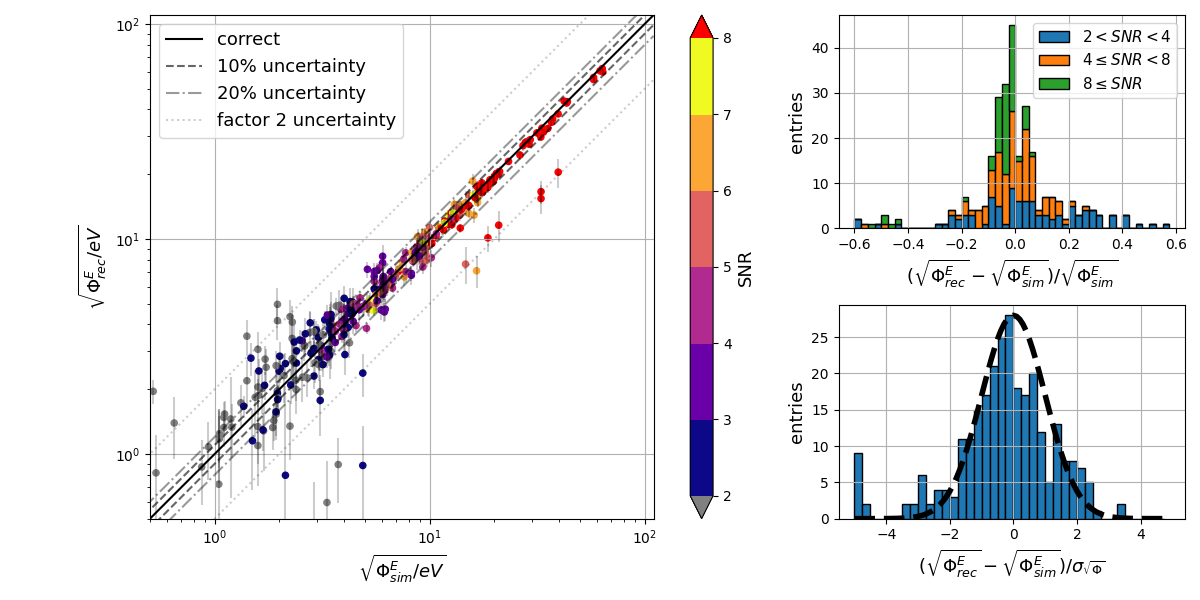}
    \caption{Reconstructed energy fluence of the radio signals from inclined air showers detected with an ARIANNA-like radio detector at the site of the Pierre Auger Observatory. Left: Scatter plot of the reconstructed square root of the energy fluence $\sqrt{\Phi^E}$ vs.\ the actual energy fluence of the radio signal. The errorbars show the uncertainty estimated by the IFT reconstruction algorithm. Top right: Stacked histogram of the errors of the reconstructed $\sqrt{\Phi^E}$ for different signal-to-noise ratios. Bottom right: Errors of the reconstructed $\sqrt{\Phi^E}$ as a multiple of the estimated uncertainty. For comparison, a Gaussian function with $\sigma=1$ is drawn as well. The left- and rightmost bins in each histogram are overflow bins.}
    \label{fig:energy_fluence_aera}
\end{figure}

\begin{figure}
    \centering
    \includegraphics[width=.9\textwidth]{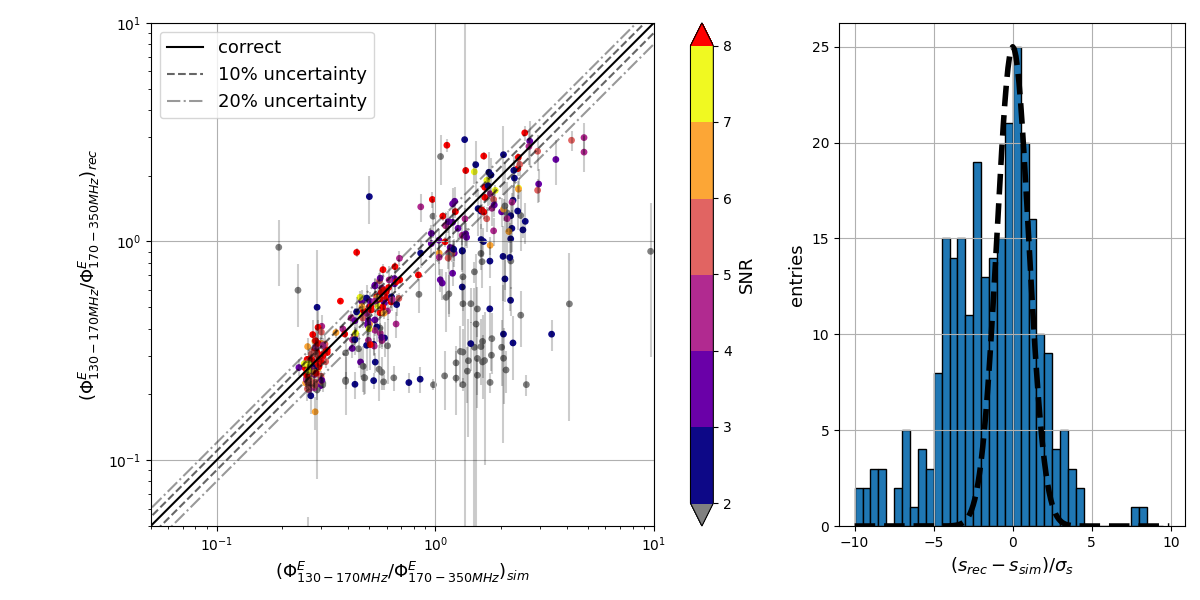}
    \caption{Reconstruction of the spectral shape parameter $s$ for radio signals from inclined air showers detected with an ARIANNA-like detector at the site of the Pierre Auger Observatory. Left: Reconstructed vs.\ true shape parameter. Error bars show the uncertainties estimated by the IFT algorithm. Right: Errors on the reconstructed shape parameter divided by its estimated uncertainties.}
    \label{fig:slope_aera}
\end{figure}

\begin{figure}
    \centering
    \includegraphics[width=.9\textwidth]{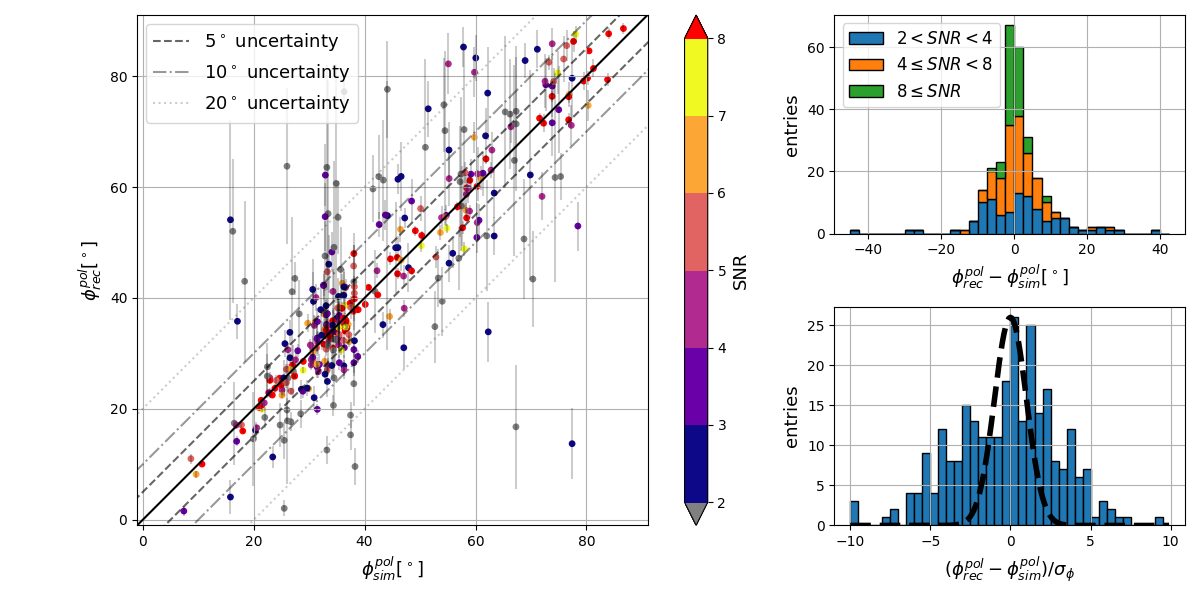}
    \caption{Reconstruction of the polarization for radio signals from inclined air showers detected with an ARIANNA-like station at the site of the Pierre Auger Observatory. Left: Reconstructed vs.\ actual polarization angle. Error bars show the uncertainties estimated by the IFT algorithm. Top right: Stacked histogram of the error on the reconstructed polarization angle for different signal-to-noise ratios. Bottom right: Errors on the reconstructed polarization angles divided by the estimated uncertainties.}
    \label{fig:polarization_aera}
\end{figure}

To assess the quality of the electric field reconstruction, we use the same three properties as for the neutrino case: The signal energy fluence, the shape parameter and the polarization.
The results for the square root of the signal energy fluence are shown in Fig.~\ref{fig:energy_fluence_mooresbay} and \ref{fig:energy_fluence_aera} for a detector at Moore's Bay and the site of the Pierre Auger Observatory, respectively. Because the LPDAs are more sensitive than the Hpol antennas used for RNO-G, especially for the $\vec{e}_\phi$ component, the energy fluence reconstruction performs better than in the neutrino case.
Because we used real noise, some events had maximum SNRs lower than 2, which are shown in gray in the scatter plot, but left out of the histograms. 

For the shape of the frequency spectrum, we use the ratio between the energy fluence in two different frequency ranges again, but we chose different passbands this time. The reason is, that the spectra of radio signals from air showers reach their maximum typically around \SI{80}{MHz}, which is outside of the recorded band, and is much weaker at higher frequencies than the signals from a shower in ice. Therefore, for many events, the energy fluence in the \SI{250}{MHz} - \SI{500}{MHz} is practically 0, making that parameter useless. Instead, we use the \SI{130}{MHz}-\SI{170}{MHz} and the \SI{170}{MHz}-\SI{250}{MHz} passbands. The results are shown in Fig.~\ref{fig:slope_mooresbay} and Fig.~\ref{fig:slope_aera}. For events with $\mathrm{SNR}>3$, the reconstruction works well, though uncertainties are again underestimated.

The results of the polarization reconstruction are shown in Fig.~\ref{fig:polarization_moorebay} and Fig.~\ref{fig:polarization_aera}. Even at SNR just over 2, the uncertainty on the reconstructed polarization angle is around \SI{10}{^\circ} and drops to a few degrees at higher SNR (see Tab.~\ref{tab:resoluton_air_shower}). Interestingly, the uncertainties are roughly the same between the simulations for Moore's Bay and the inclined showers at the Auger site, despite the low sensitivity to $\vec{e}_\theta$ for inclined showers.

\begin{table}
\centering
\begin{tabular}{|c|c|c|c|}
     \hline
     &\multicolumn{3}{c|}{Moore's Bay} \\
     & $2 \leq \mathrm{SNR} < 4$ & $4 \leq \mathrm{SNR} < 8$ & $\mathrm{SNR} < 8$\\
     \hline
     $\Delta \Phi^E$ &&&  \\
     68\% qtl. & [-0.17, 0.62] & [-0.13, 0.12]& [-0.11, -0.02]\\
     median &0.20 & -0.01 & -0.07\\
     \hline
     $\Delta \sqrt{\Phi^E}$ &&&\\
     68\% qtl.& [-0.08, 0.28]& [-0.07, 0.06] & [-0.05, -0.01]\\
     median & 0.10 & -0.01 & -0.04\\
     \hline
     $\Delta \phi_\mathrm{pol}$ &&&\\
     68\% qtl. &[\SI{-17.3}{^\circ}, \SI{3.2}{^\circ}] & [\SI{-8.6}{^\circ}, \SI{3.61}{^\circ}] & [\SI{-1.5}{^\circ}, \SI{2.7}{^\circ}]\\
     median &\SI{-2.8}{^\circ} & \SI{0.0}{^\circ} & \SI{0.4}{^\circ}\\
       \hline \hline
     & \multicolumn{3}{c|}{Auger}\\
      & $2 \leq \mathrm{SNR} < 4$ & $4 \leq \mathrm{SNR} < 8$ & $\mathrm{SNR} < 8$\\
     \hline
     $\Delta \Phi^E$ &&&  \\
      68\% qtl. & [-0.29, 0.62] & [-0.17, 0.12] & [-0.13, -0.01] \\
      median &0.07 & -0.03 & -0.07\\
      \hline
      $\Delta \sqrt{\Phi^E}$ &&&\\
      68\% qtl.& [-0.15, 0.27] & [-0.09, 0.06] & [-0.07, 0.00] \\
      median &0.04 & -0.01 & -0.03 \\
      \hline
      $\Delta \phi_\mathrm{pol}$ &&&\\
      68\% qtl. &[\SI{-8.0}{^\circ}, \SI{11.0}{^\circ}] & [\SI{-3.8}{^\circ}, \SI{5.7}{^\circ}] & [\SI{-1.5}{^\circ}, \SI{2.0}{^\circ}] \\
      median &\SI{1.1}{^\circ} & \SI{0.2}{^\circ} & \SI{-0.3}{^\circ} \\
     \hline
\end{tabular}
\label{tab:resoluton_air_shower}
\caption{Uncertainties on the reconstruction of the signal energy fluence $\Phi^E$, its square root and the polarization angle $\phi_\mathrm{pol}$ for radio signals from air showers detected by an ARIANNA station at Moore's Bay and from inclined air showers detected by an ARIANNA station at the site of the Pierre Auger Observatory. For each quantity, the 68\% quantiles and the median of the distribution of errors in the reconstructions are given. For $\Phi^E$ and $\sqrt{\Phi^E}$ they are relative errors, fpr $\phi_\mathrm{pol}$ they are absolute errors, in degrees.}
\end{table}

\subsection{Implications for radio detectors}
\label{sec:implications}
We have shown that \emph{Information Field Theory} can be used to reconstruct radio signals detected with the Radio Neutrino Observatory in Greenland (RNO-G), even at low signal-to-noise ratios just above the trigger threshold.

The energy fluence of the radio signal is proportional to the square of the shower energy, so the relative uncertainty on $\sqrt{\Phi^E}$ translates directly to the uncertainty on the shower energy. However, the uncertainty on the fraction of the neutrino energy that is deposited into the shower is around a factor of 2 \cite{Anker:2019zcx}, thus dominating over the uncertainty on the energy fluence. Therefore, further improving the energy fluence reconstruction will have very little effect on the reconstructed neutrino energies.

The shape of the electric field spectrum affects both the energy and the neutrino direction reconstruction, as it depends on the angle to the shower axis at which the signal is emitted. As shown in Fig.~\ref{fig:signal_examples}, this relationship is difficult to quantify in general and to attempt this beyond the scope of this paper. As we are not aware of another robust method to obtain the frequency spectrum at low SNR events, other than fitting a specific electric field model to the data \cite{NuRadioReco}, IFT currently seems like the best option to do this for neutrino detection.

The resolution on the signal polarization is mainly limited by the low gain of the horizontally-polarized antennas resulting from their constrained geometry. The polarization is needed to constrain the neutrino direction: If one only knows the direction of the radio signal, the neutrino direction can only be constrained to a cone with an opening angle equal to the emission angle, which is close to the Cherenkov angle. Using the fact that the radio signal is polarized radially around the shower axis, a measurement of the polarization can be used to resolve where on the cone the neutrino direction is. Thus, if the polarization angle is misreconstructed by a small angle $\Delta \phi_\mathrm{pol}$ it causes the neutrino direction to be off by
\begin{equation}
    \delta \approx \sin(\theta_C) \cdot \Delta \phi_\mathrm{pol} \approx 0.8 \cdot \Delta \phi_\mathrm{pol}
\end{equation}
with a Cherenkov angle $\theta_C=56^\circ$.

While these IFT results already look promising, it should be noted that only six out of the 24 channels available for RNO-G were used in this study. This was done because the IFT model used required the radio signal at all used channels to have a very similar shape and roughly the same amplitude, which can be assumed to be always the case for the channels in the phased array and the Vpol antennas directly above them, but not necessarily for the others. How much the radio signal varies between channels depends on the event geometry, most importantly the distance to the shower. So if the event geometry is known, one can determine if it makes sense to include additional channels. This decision can also depend on the signal-to-noise ratio and the goal of the reconstruction: If some channels have a very high SNR, it may make sense to reconstruct each of them individually to compare the radio signal at different positions, while at low SNR, all available channels may be necessary to get a estimate of the signal properties. While these considerations are beyond the scope of this paper, they can lead to further improvements in the signal reconstruction.
Finally, additional channels could be included by dropping the requirement that the spectral shape has to be identical from the IFT model. IFT offers options to build models in which the shape of the spectrum is correlated between channels, but does not have to be the same.

As in the neutrino case, the energy of cosmic-ray induced air showers is proportional to the square root of the energy fluence of the radio signal. While the uncertainty on the signal energy fluence for a detector at Moore's Bay is comparable to the one obtained with the standard reconstruction \cite{NuRadioReco}, IFT performs much better regarding the polarization, with a similar accuracy to the \emph{forward folding} method.

To our knowledge, the only radio detector currently using the shape of the frequency spectrum for cosmic ray detection is ANITA \cite{ANITA_first_flight}, where a parametrization was fitted to the electric field spectrum of the radio signal in order to estimate the energy. As the electric field was reconstructed by unfolding the measured voltages and the antenna responses, this approach requires a large signal-to-noise ratio, so IFT could be used in order to lower the reconstruction threshold.
A similar approach has been developed for ground-based cosmic ray detectors \cite{Welling:2019scz}, in which the \emph{forward folding} technique was used to reconstruct the radio signal. While IFT would likely not improve the precision of the energy reconstruction, not having to rely on a specific parametrization can be an advantage in itself, if one wants to use a lower passband where the parametrization with an exponential function no longer holds or if the origin of the radio signal may not be entirely clear.
Unfortunately, most radio detectors have to limit their band to the \SIrange{30}{80}{MHz} band, because of the noise from radio communications at other frequencies. This small band limits the use of the spectrum for event reconstruction.

When reconstructing the simulated data set for the Pierre Auger Observatory, the IFT method showed a similar performance, demonstrating that it also works for inclined showers. Currently, the Pierre Auger Observatory only uses the horizontal components of the electric field if the incoming zenith angle is larger than \SI{60}{^\circ} \cite{aera_inclined_showers}, meaning that information about the full radio signal is lost and the signal polarization is practically unknown. With IFT, the full electric field could be made available, which should improve the performance of the event reconstruction. 

\subsection{Cross-check: Electric field reconstructions on pure noise}
\begin{figure}
    \centering
    \includegraphics[width=.9\textwidth]{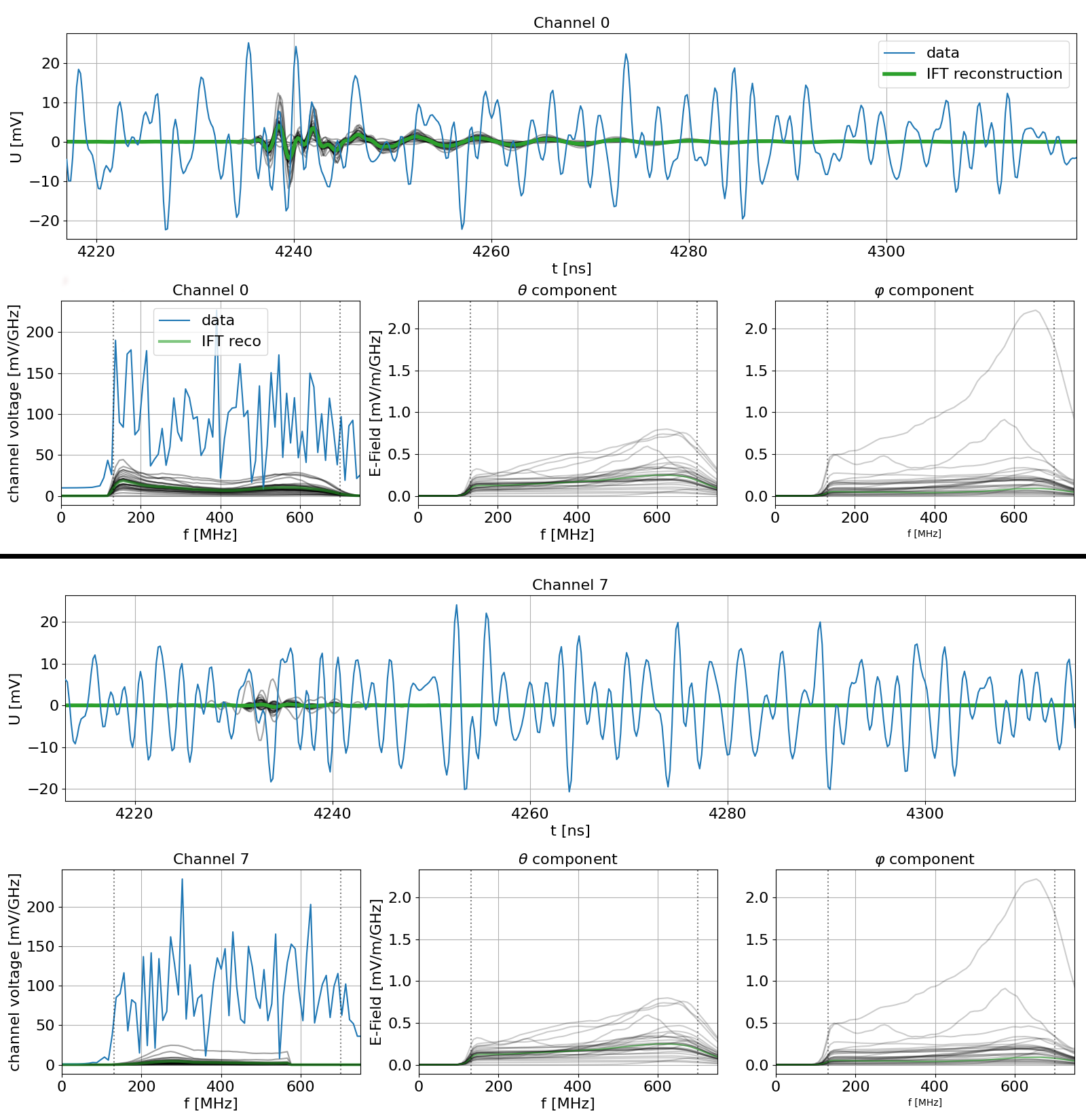}
    \caption{Results of the reconstructed electric field of an RNO-G event containing only noise, for one of the Vpol (top) and Hpol (bottom) channels. 1st and 3rd row: Waveforms of the data (blue) and reconstructed voltage (green). 2nd and 4th row: Spectrum of the voltage (left) and the $\vec{e}_\theta$ (center) and $\vec{e}_\phi$ component of the electric field. The thin black lines are the samples used to estimate uncertainties, which are drawn to visualize the uncertainties on the reconstruction.}
    \label{fig:noise_reco}
\end{figure}
If a detector is self-triggering, there is a chance that a random noise fluctuation is mistaken for a genuine trigger. In this case, it is desirable that the electric field reconstruction is not forced to return a realistic-looking result, but either provides evidence that something is wrong or returns a result that looks non-physical to make it clear that no signal was present.

To test how IFT behaves, we run the same reconstruction we used for RNO-G events, but this time the channels recorded pure noise. A typical example of the result of such a reconstruction is shown in Fig.~\ref{fig:noise_reco}. The reconstructed electric field is very small, so that the reconstructed voltage is very small as well, compared to the noise level. This alone is evidence that there is likely no actual radio pulse in the data, but looking at the thin gray lines visualizing the uncertainties, we see that the electric field being 0 is within the margin of error. Thus, this result can be considered a ``correct" electric field reconstruction, as it returns that the radio signal is, at most, very small.

\subsection{Cross-check: Systematic uncertainties}


\begin{figure}
    \centering
    \includegraphics[width=.9\textwidth]{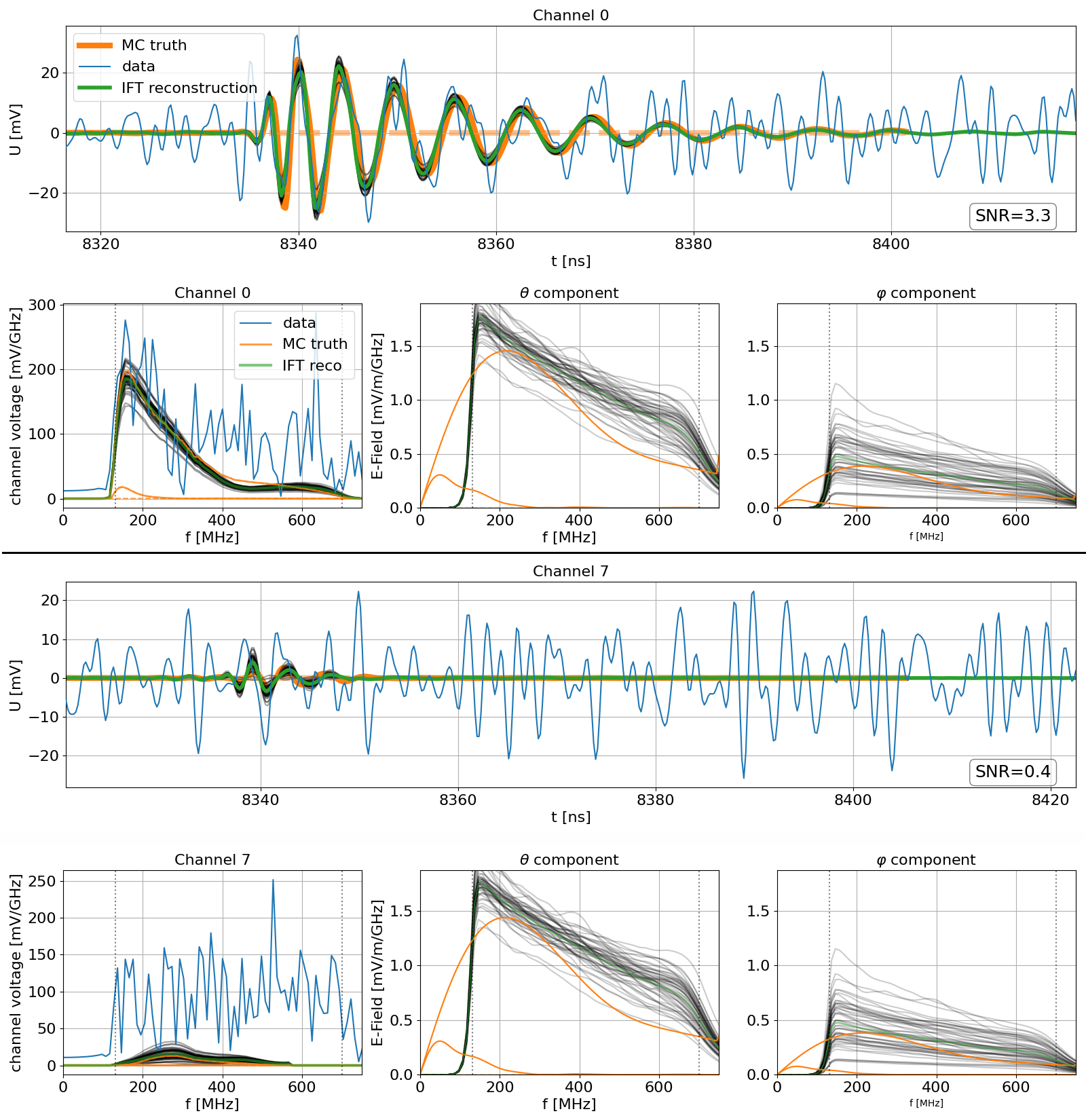}
    \caption{Result of the electric field reconstruction for the same event as shown in Fig.~\ref{fig:reco_example}, but for this an incorrect Vpol model is used for the reconstruction.}
    \label{fig:reco_50cm_vpol}
\end{figure}

So far we used the same detector response for the simulations and the reconstruction, which implies perfect knowledge of the detector properties. In reality, this is of course hardly possible. While the response of most parts of the signal chain can be measured with great precision in the lab prior to deployment, modelling of the antennas carries larger uncertainties. The surrounding ice, for example, influences the antenna response, which is difficult to recreate in the lab. Therefore the most common way to solve this problem is to use simulations and cross check them with pulser measurements after deployment.

In addition, the antenna gain and phase response also depends on the signal arrival direction at the antenna, so a incorrect reconstruction of the signal direction will also result in using an incorrect antenna response.

For RNO-G several revision of Vpol antennas and bore-hole diameters were considered and fully simulated. 
We can make use of this to mimic the effect of an incorrect antenna model by running the IFT reconstruction on the same event as shown in Fig. \ref{fig:reco_example}, except this time the antenna model for an earlier revision is used. This represents a very pessimistic assumption about the uncertainty on the antenna model, as the for the revision included simulations were run with a borehole almost twice as wide, resulting in both a different overall shape and resonances in the model.

The result is shown in Fig.~\ref{fig:reco_50cm_vpol}. At first sight, it looks like the reconstruction of the electric field performed rather poorly. However, voltages measured in the channels are still reconstructed well. Comparing the two antenna models in detail, it explains that the electric field amplitudes at frequencies below \SI{200}{MHz} and above \SI{400}{MHz} are overestimated, since the used Vpol model is much less sensitive there.

From this illustrative exercise, we can conclude that a systematic uncertainty on the antenna response it not an issue for the IFT method itself, but will influence how the voltages measured by the channels are translated into an electric field.

\section{Conclusion}
One of the key steps in the event reconstruction for radio-based particle detectors is determining the electric field of the radio signal from the voltage waveforms measured with its antennas, which is difficult to do accurately because they are always contaminated by noise from the surrounding environment. The \emph{Information Field Theory} method presented in this paper offers a solution to this issue. 

When applied to a simulated neutrino event, IFT moderately improves the reconstruction of the overall electric field amplitude, but its main strength lies in the ability to reconstruct the shape of the frequency spectrum and the polarization, even at low signal-to-noise ratios. This is especially important for neutrino detectors, where most events are expected to be close to the trigger threshold, and the low event rate prohibits strict quality cuts. Experiments like the Radio Neutrino Observatory Greenland (RNO-G) or the radio component of IceCube-Gen2 will certainly profit from applying IFT to the detected pulses. By using a non-parametric model for the frequency spectrum, we make minimal prior assumptions about the expected signal, which is an advantage for an experiment attempting the first detection of a neutrino-induced particle shower via its radio signal.

Making minimal assumptions about the radio signal also has the advantage that the same method can be applied to other kinds of radio pulses, like those emitted by an air shower. The improved electric field measurement allows new reconstruction methods using the shape of the frequency spectrum and the polarization of the radio signal. With IFT, it is also possible to reconstruct the full electric field for inclined air showers.

For these reasons, Information Field Theory can be a valuable asset to improve the precision of radio-based particle detectors and move their analysis threshold to lower energies.

\section{Acknowledgements}
We are grateful to the ARIANNA collaboration for allowing us to use a fraction of their noise data to show the suitability of the method for measured data. 
This work was carried out with support from the German research foundation (DFG), under grant NE 2031/2-1.

The source code to perform the reconstructions shown in this paper is publicly available as part of the \emph{NuRadioReco} open source software package \cite{NuRadioReco}.

\bibliographystyle{bibstyle}
\bibliography{references}
\appendix
\section{Failed Reconstructions}
\label{sec:failed_recos}
\begin{figure}
    \centering
    \includegraphics[width=.9\textwidth]{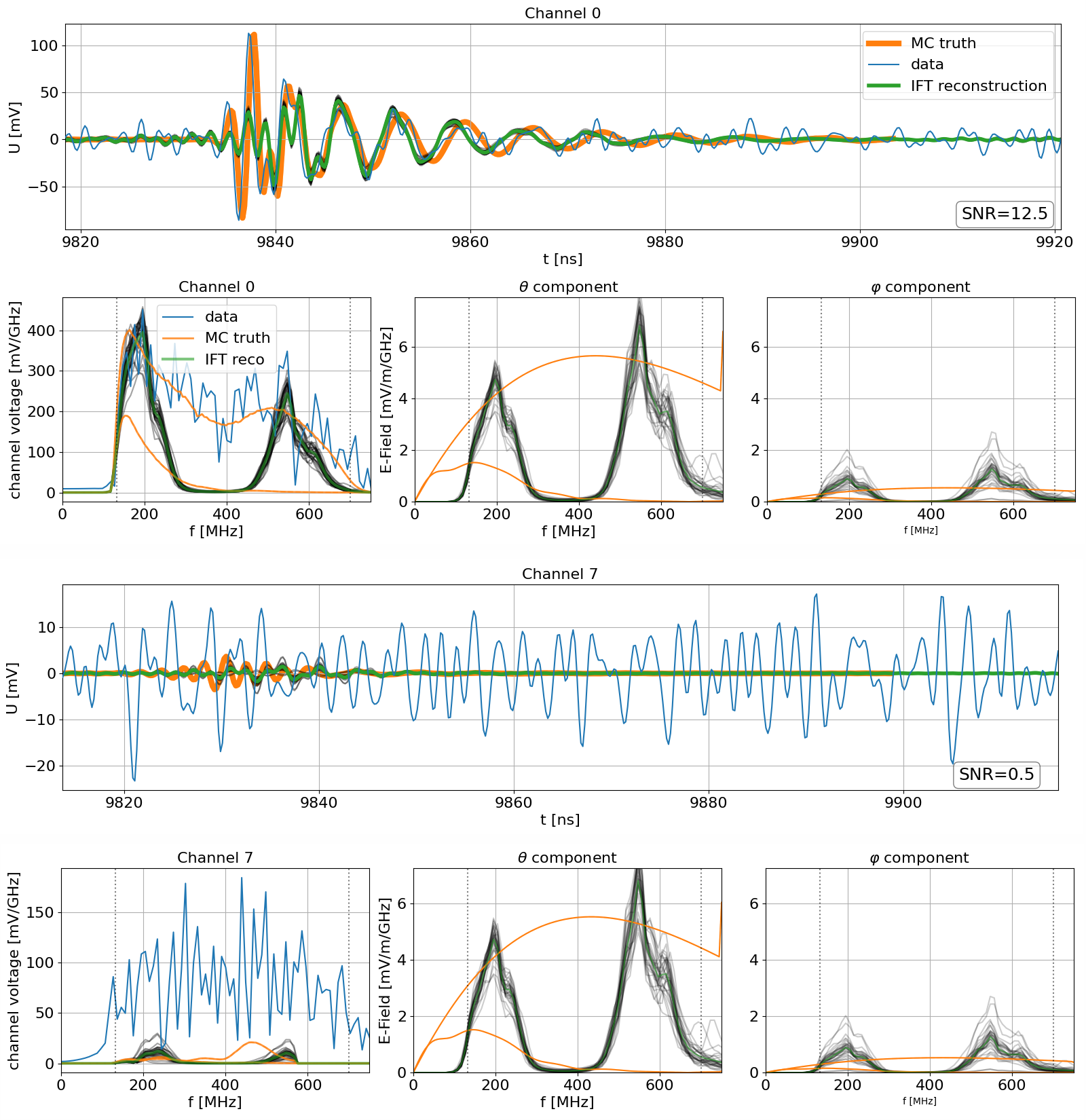}
    \caption{Results of a failed electric field reconstruction of an event detected by an RNO-G station. Shown are one of the Vpol (top) and one of the Hpol (bottom) channels. 1st and 3rd row: Waveform of the data (blue), the actual signal (orange) and the reconstruction (green). 2nd and 4th row: Spectrum of the channel voltage (left) and the $\vec{e}_\theta$ (center) and $\vec{e}_\phi$ (right) component of the electric field. The thin gray lines are the samples used for uncertainty estimation.}
    \label{fig:failred_reco}
\end{figure}
As shown in, for example, Fig.~\ref{fig:fig:energy_fluence_rno}, sometimes the IFT reconstruction produces outliers, where the result does not come close to the actual radio signal. These outliers occur mainly for two reasons:

In the case of the neutrino detector, there can be events that do not match the IFT model. The most common cause is when two different radio pulses arrive shortly after another and overlap. This can happen with certain event geometries, in which the propagation time differences between the ray that reaches the antenna directly and the one that is refracted downwards are very small. 

The other way a double pulse can occur is if a high-energy electron neutrino interacts via charged current interaction. This produces two showers: One hadronic shower from the nucleon recoil and one electromagnetic shower from the electron produced in the process. Usually, both showers overlap and their radio signals can be treated as coming from a single shower. However, at high energies the LPM effect \cite{lpm1, lpm2} elongates the electromagnetic shower \cite{PhysRevD.82.074017}. As the particle showers propagate through the ice faster than the radio signals, there can be a noticeable time difference between the signal from the hadronic and the electromagnetic showers. In that case, there are practically two different radio pulses.

As we assume that only one radio signal is present in the data, the model is thus not suitable those two situations. 


The second reason for the IFT reconstruction to fail is that the variational inference can converge on a local minimum, if the slope of the phase is estimated incorrectly. This can cause the IFT reconstruction to estimate a wrong signal timing, and only fit to part of the radio pulse. This is especially a problem for detectors where a large group delay in the signal chain elongates the signal recorded by the channels. This is the case for the current ARIANNA detector and argues for a signal chain with less group delay for future detectors.

For in-ice showers, a second effect plays a role: Because the shower front propagates significantly faster than radio signal, signals from the early stage of the shower development arrive after the ones from the later stages, if the observer is located inside the Cherenkov ring. To such an observer, it looks like the shower propagation is happening in reverse, and the radio signal is mirrored on the time axis. In Fourier space, this corresponds to a sign change in the slope of the linear function describing the phase, compared to an observer outside the Cherenkov ring. The prior probability distribution of the slope is centered around one value, so only one of these cases (the observer outside the Cherenkov ring in this paper) is taken into account. Fortunately, most of the phase of the measured voltage is dictated by the antenna and amplifier response, so the reconstructions usually work for observers inside the Cherenkov ring as well. Nevertheless, the reconstruction quality could be improved by testing both configurations of the phase slope and determining which one yields a better result. An additional benefit would be the ability to determine which side of the Cherenkov ring the antenna is on, which provides additional insight on the event geometry.

Fortunately, both of these mis-reconstructions are easily identified. A typical example is shown in Fig.~\ref{fig:failred_reco}: The reconstruction converged on a pulse timing that misses the start of the radio signal and tries to find a solution that matches the tail of the pulse. However, much of pulse shape is dictated by the group delay of the antenna and amplifier, so it is difficult to find a reconstruction for the amplitude of the frequency spectrum that matches the data. The result is a spectrum that has multiple peaks and drops to zero in between. This feature in the spectrum can be used to identify events where the electric field reconstruction likely failed.

Of course, assuming that the signal timing is known, it is also possible to check if the signal time of the reconstructed radio pulse matches.

\end{document}